\documentclass[12pt,preprint]{aastex}

\newcommand{\eb}{\begin{equation}}
\newcommand{\ee}{\end{equation}}

\def\3dot#1{\stackrel{...}{#1}}

\shorttitle{Astrometric orbits}
\shortauthors{Goldin \& Makarov}

\begin{document}

\title{Unconstrained Astrometric Orbits for Hipparcos Stars with Stochastic Solutions}

\author{A. Goldin \altaffilmark{1} \& V.V. Makarov \altaffilmark{2}}
\affil{$^1$ Citadel Investment Group, 131 South Dearborn st., Chicago, IL 60603}
\email{alexey.goldin@gmail.com}
\affil{$^2$Michelson Science Center, Caltech, 770 S. Wilson Ave.,
MS 100-22, Pasadena, CA 91125}
\email{vvm@caltech.edu}

\begin{abstract}
A considerable number of astrometric binaries whose positions on the sky
do not obey the standard model of mean position, parallax and linear proper motion,
were observed by the Hipparcos satellite. Some of them remain non-discovered,
and their observational data have not been properly processed with
the more adequate astrometric model that includes nonlinear orbital motion.
We develop an automated algorithm based on ``genetic
optimization'', to solve the orbital fitting problem in the most difficult setup, when no prior information
about the orbital elements is available (from, e.g., spectroscopic data or radial velocity
monitoring). We also offer a technique to accurately compute the probability that an
orbital fit is bogus, that is, that an orbital solution is obtained for a single star, and to estimate the probability
distributions for the fitting orbital parameters. We test this method on Hipparcos stars
with known orbital solutions in the catalog, and further apply it to 1561  stars with stochastic
solutions, which may be unresolved binaries.  At a confidence level of 99\%,
orbital fits are obtained for 65 stars, most of which have not been known as binary.
It is found that reliable astrometric
fits can be obtained even if the period is somewhat longer than the time span of
the Hipparcos mission, i.e., if the orbit is not closed. A few of the new probable binaries with A-type primaries with
periods 444--2015 d are chemically peculiar stars, including Ap and $\lambda$ Boo type. The anomalous
spectra of these stars are explained as admixture of the light from the unresolved, sufficiently
bright and massive companions. We estimate the apparent orbits of four stars which have been
identified as members of the $\approx 300$ Myr-old UMa kinematic group. Another four new nearby binaries
may include low-mass M-type or brow dwarf companions. Follow-up spectroscopic observations in conjunction with more
accurate inclination estimates will lead to better estimates of
the secondary mass. Similar astrometric models and algorithms can be used
for binary stars and planet hosts observed by SIM PlanetQuest and Gaia.

\end{abstract}


\section{Introduction}

One of the most promising and presently available ways to discover invisible companions (including
brown dwarfs and giant planets) to nearby Galactic stars is to analyze high-accuracy
astrometric positions of the latter for the presence of the reflex
Keplerian motion caused by an orbiting companion. In the relatively near
future, with the advance of such space-borne astrometric instruments
as SIM and GAIA, it will become one the main instruments in the search for
habitable smaller planets, too. For the time being, the
capabilities of the method are limited by the moderate precision
of the available astrometric data. The Hipparcos Intermediate Astrometry
Data (HIAD), at a single point precision of roughly 10--15 mas, is just good enough
to detect brown dwarfs around nearby solar-type stars, reliably so
in conjuction with spectroscopic measurements \citep{ha,pou,tor}. Perhaps,
giant planets of intermediate orbital periods could
also be detected around a few nearest stars.

Obtaining a robust orbital fit for an astrometric binary becomes
increasingly difficult when the orbital period $P$ exceeds the time span
of Hipparcos measurements, which is about 3.2 years. If the period is
6 years, still more than half of the orbit is represented in the
data, and astrometric analysis may provide an independent solution.
When the period is longer than 9 years, the astrometric data alone
can provide only ambiguous results, since the almost linear segment
of the orbit is hard to distinguish from the regular proper motion.
Generally, the astrometric orbital fit is a non-linear 12-parameter adjustment
problem that includes five astrometric parameters and seven orbital parameters
(Appendix A). In some cases, e.g., long-period orbits, or highly inclined
orbits aligned with the line of sight, the parameters become so entangled
in the non-linear condition equations, that the fit becomes ill-conditioned.
Therefore, any orbital solution should be supported by analysis of confidence intervals
for all fitting parameters. If the astrometric solution is sufficiently well-constrained,
approximate standard deviations of fitting parameters can be derived in the near vicinity of
the solution point by covariance analysis, a recipe of which is given in Appendix B. 
Ill-conditioned solutions will have nearly singular normal matrices that can not be inverted, or large
variances for some of the parameters.
In more complicated cases, some of which we investigate in this paper,
the confidence intervals and probability distributions can be calculated by
extensive Monte-Carlo simulations. When the optimization problem
is almost singular with respect to particular nonlinear parameters (most often the
eccentricity), direct mapping of the objective function on a sufficiently fine grid of parameters
is carried out.

Another crucial problem that arises in fitting Keplerian motion for stars previously not known as
binaries, is estimation of the probability of a null hypothesis, that is, that the star is single, and the
detected perturbation in the astrometric residuals is caused by a chance occurrence of
random noise in the data, or other effects, unrelated to binarity. We offer a robust and straightforward
(if somewhat computationally heavy) method of confidence level estimation of a binary
detection.

These techniques are tested on a set of 1561 stars in the Hipparcos catalog with the so-called stochastic
solutions. These data often represent gross errors or failures in the data reduction. Due to the type of
detector used in the main instrument of Hipparcos (a grid of long vertical slits and a photomultiplier behind it
with a non-uniform response across the pointing field of regard), many of these failed solutions originate
in the wrong assumption about the target positions, or in the lack of knowledge of the relative position and
brightness of visual binary components. Successful attempts have been made to use more accurate
information about stochastic stars from ground based and Tycho-2 data sets to reprocess the published
Hipparcos transit data, resulting in accurate solutions for a few hundred stars \citep{fal,fama}.
The remaining stars with stochastic solutions are prime suspects for yet unknown binaries. Orbiting pairs
with separations smaller than $0.1$ arcsec were not resolved by Hipparcos; on the other hand, a nonlinear
apparent motion of the photocenter as small as several milliarcseconds could break the regular 5-parameter solution.
Long-period orbits could be detected as additional acceleration of proper motions or as discrepant proper
motions from the short-term Hipparcos data and the long-term Tycho-2 definitions \citep{maka}. The overlap
between such accelerating astrometric binaries and the list of stochastic objects is substantial, but presumably
a lot of binaries on shorter orbits are left out of this analysis, as they describe a strongly nonlinear track during
the 3.2-3.5 time span of Hipparcos observations.

\section{Using HIAD to compute orbits}

\subsection{Correlated input data}
The star abscissae data in the HIAD was derived by the two Hipparcos data reduction
consortia, FAST and NDAC, almost independently, but from the same observational
data \citep[Vol.~3]{esa}. Typically, each great circle measurement produced
a pair of abscissa data points, one derived by FAST, and the other by NDAC.
These pairs of measurements are statistically correlated. The correlation
coefficient was estimated and published in the HIAD record, when
appropriate. However, some of the great circle results were processed
(or accepted) by only one of the consortia, in which case the correlation
is zero. Since any astrometric or orbital fit is a least-squares
adjustment minimizing the $\chi^2$ on abscissa residuals, internal
correlations must be properly taken into account. 

In this paper, we obtain separate orbital fits based on NDAC and FAST data. In our judgment,
these data are derived from the same observations with the same instrument, and they
should be strongly correlated. The correlations given in the HIAD are probably strongly
underestimated, and using a weighted combination of these data may have an adverse effect
on the confidence estimation.

\subsection{Multiple solutions and local minima}

The orbital motion model described in Appendix \ref{A1.sec} contains 3
nonlinear and 9 linear parameters. After $P_0$, $e$ and $T_0$ are
fixed by a nonlinear optimization method, the rest of parameters ($A$,
$B$, $F$, $G$, and 5 astrometric parameters) can be found by solving a
linear system of equations.

Solving a multidimensional optimization problem is difficult.
Deterministic methods such as the widely used Powell minimization on
starting conditions can converge to different local minima.
Two-dimensional optimization is sometimes marginally solvable by brute
force, i.e., by walking over a grid of initial values and initiating
optimization from each of these starting values. For 3 and more
dimensions, however, this approach is time consuming and a better
approach is welcome (Section 6).

\section{Genetic optimization algorithm}

We employ a method described in \cite{DE95}, which is a generalization
of the genetic algorithms for continuous functions optimization.
The differential evolution (DE) algorithm is capable of handling nondifferentiable,
nonlinear objective functions. It is one of the methods used in {\em
  Mathematica} \verb|NMinimize| routine. Although our $\chi^2$ objective function
is differentiable everywhere except several singular points (Appendix B) and, with respect to a few parameters, linear,
we have chosen this algorithm for its enhanced ability to handle complex surfaces
in the parameter space with multiple minima, as discussed in Section \ref{conf.sec}.

In a population of potential solutions in a $n$-dimensional search
space, a fixed number of vectors are randomly initialized, then
evolved over time to explore the search space and to locate the minima
of the objective function.

At each iteration, called a generation, new difference vectors are
computed by subtracting two vectors selected randomly from the
current population (mutation). The difference vector is added to a
third vector, which may be the current optimal vector or a randomly
selected one.  The resulting candidate is mixed with the current best
vector, imitating mixing of chromosomes in sexual reproduction, where
each coordinate component corresponds to a chromosome.  The resulting
trial vector is accepted for the next generation if it yields a
reduction in the value of the objective function. This process,
imitating biological evolution, is not guaranteed to find the best
solution that corresponds to the global minimum of the objective function, just
as evolution does not always select the best possible mutation. However,
the probability of finding the global minimum, or a very close solution to the
global minimum is increasing with the population size. Therefore, at the expense
of extra computing time, we can find the `nearly' best solution within a given model,
localizing most of the local minima (alternative solutions) at the same time by
assuming large numbers of generation vectors. This allows us to investigate the multitude
of alternative solutions, rather than rely on a single possible solution that yields
a sufficiently small $\chi^2$. 

Existence of multiple solutions with roughly equally reduced $\chi^2$ is a warning
that the pattern of noisy data is untenable for a reliable binary solution, or that
some other non-binarity effect is responsible for the observed perturbation.

\section{Distributions, probabilities and confidence intervals}
\label{conf.sec}

Instead of the standard F-test, unsuitable for strongly nonlinear
models, the following Monte-Carlo test is used to evaluate the
robustness of our orbital solutions.  We generate a set of observations
of a star with coordinates, parallax and proper motions corresponding
to the best fit 5-parameter model at the same observation times.
Normally distributed random numbers with variances corresponding to
the estimated (formal) measurements' errors are added to each recorded
transit time. After that the realization is reduced for both
5-parameter and 12-parameter models, resulting in $\chi^2_{5_i}$ and
$\chi^2_{{12}_i}$ for $i=1,2,\ldots,N$, respectively.  The resulting fits are
compared with the original, unperturbed fits with corresponding
$\chi^2_{{12}_0}$ and $\chi^2_{5_0}$ statistics.
 
We estimate the fraction of trials for which the ratio of the 5-parameter fit to that of the 12-parameter fit
exceeds that ratio for the actual data:
\begin{equation}
  \label{eq:nonl-f-test}
  p={1\over N}\sum\limits_{i=1}^N \Theta_i\left(\chi^2_{5_i}/\chi^2_{12_i} -\chi^2_{5_0}/\chi^2_{{12}_0}\right)
\end{equation}
where $\Theta(x)$ is a threshold function, equal to 1 for $x>0$, and $0$ for
$x<0$. The confidence of rejecting a null hypothesis is $1-p$.  This
test is numerically equivalent to the $F$-test for a linear problem at
sufficiently high $N$.

If the probability test is pssed, extensive Monte-Carlo simulations are carried out to estimate the errors for
each parameter. Each estimated transit time is perturbed by a normally
distributed random number with a standard deviation equal to the formal
transit time error (given in the HIAD) and the reduction process is repeated.  This produces a
distribution of true transit times given observation, provided we have
accurate transit time error estimates, and the errors are truly
Gaussian. By reducing these data, we obtain the distribution of underlying
parameters given observation. This method of parameter error estimation is known as  
parametric bootstrap \citep{Statlearn}.

At least 1000 trials are performed for each object. From the collected
Monte-Carlo data, we build histograms for each fitting parameter such
as $e$, $a_a$, $T_0$, which are subsequently used to estimate
$99$, $95$ and $68$ percent confidence intervals. The distribution of
solutions is often significantly different from a Gaussian
bell curve. With increasing number of measurements, and decreasing relative
errors of parameters, the parameter likelihood function becomes confined to an area where
linear approximation is quite adequate for error estimation (Appendix B).

\begin{figure}[htbp]
  \centering
   \includegraphics[angle=0,width=0.9\textwidth]{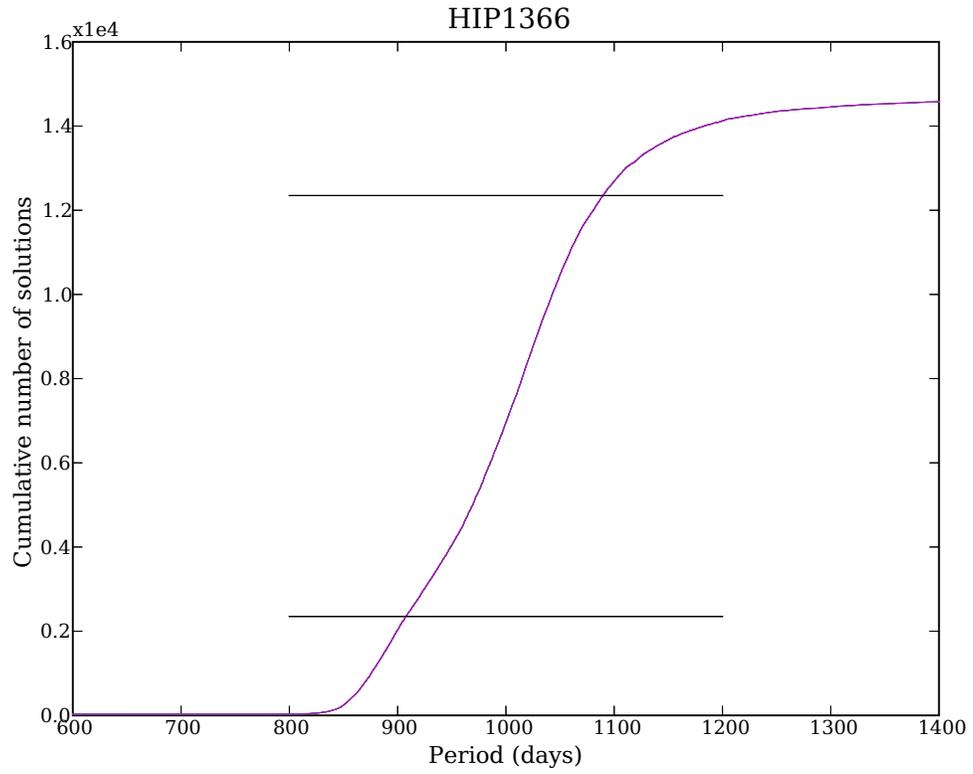}
   \caption{Period cumulative distribution for HIP1366. The total
     number of trials is 14700. Horizontal lines show bounds of 68\%
     confidence interval.}
  \label{fig:hip1366_period}
\end{figure}

\begin{figure}[htbp]
  \centering
  \includegraphics[angle=0,width=0.45\textwidth]{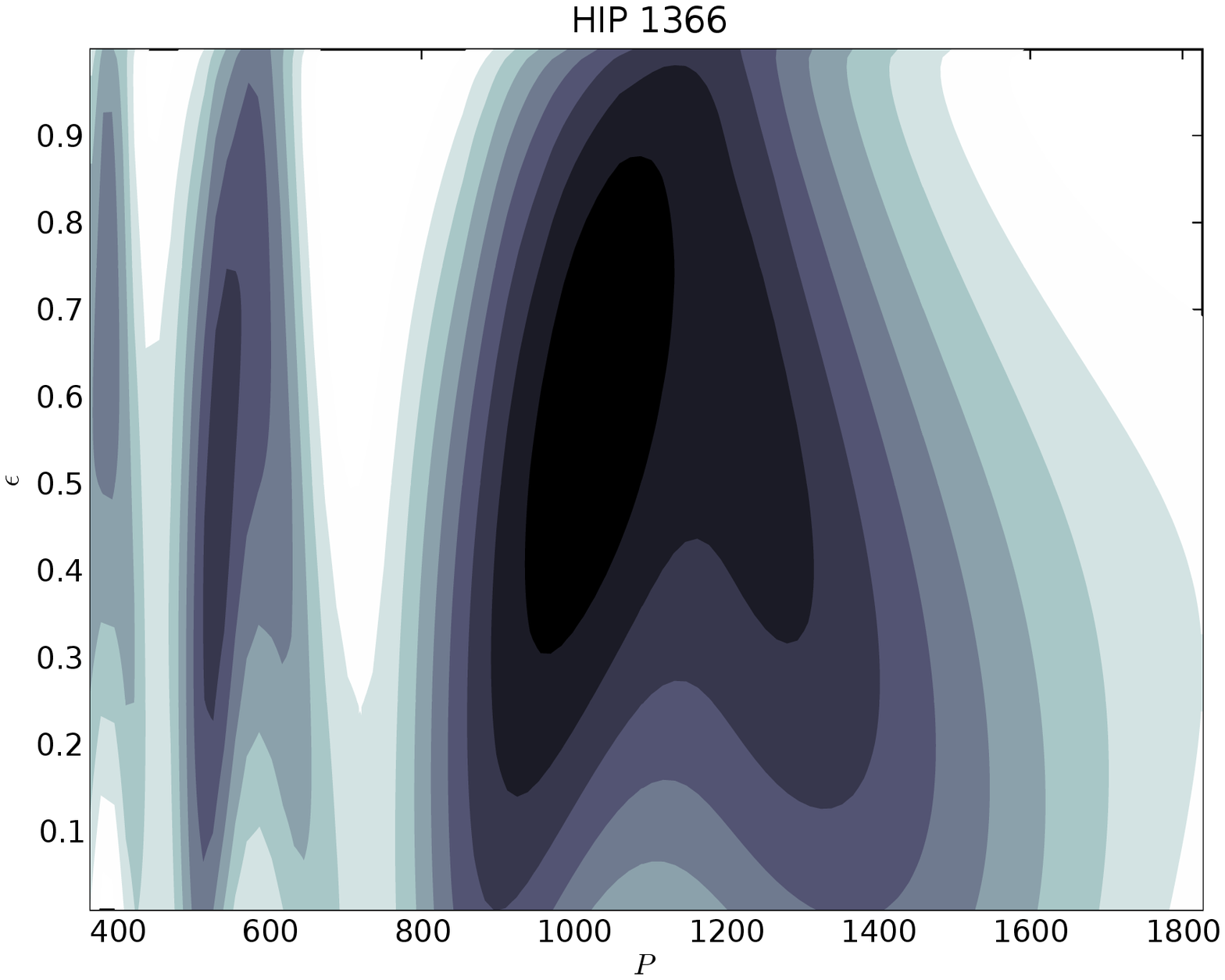}
  \includegraphics[angle=0,width=0.45\textwidth]{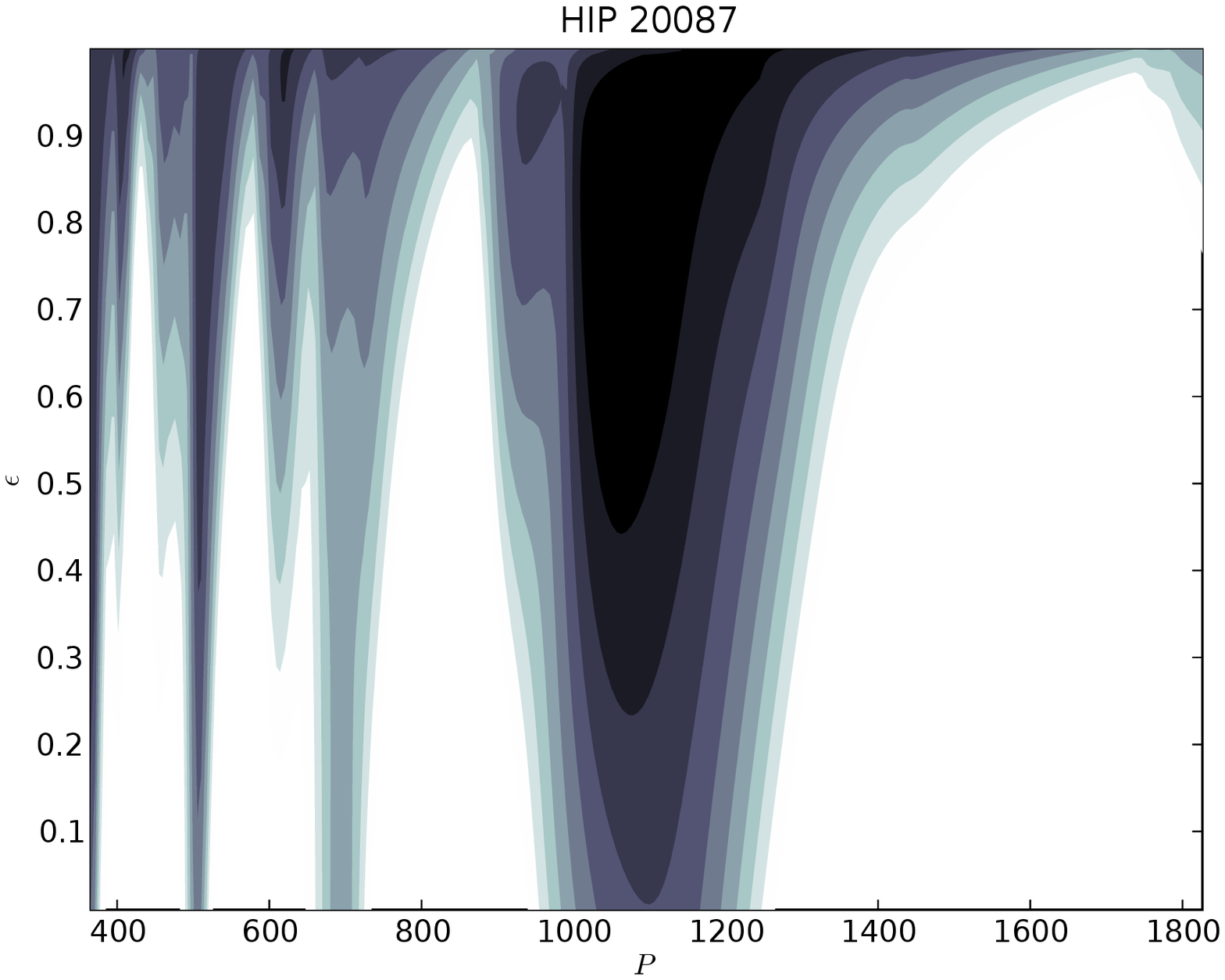}
  \caption{$\chi^2$ contours for two stars as functions of period and eccentricity.}
  \label{fig:contours}
\end{figure}

Figure \ref{fig:contours} shows $\chi^2$ computed for two representative stars, as functions of period $P$
in days and eccentricity $e$, while keeping $T_0$ and the
remaining 9 parameters constant. The spacing between $\chi^2$ contours
is 50, the darkest color corresponding to the smallest $\chi^2$. We find a well
defined global minimum for HIP 1366, while for HIP 20087 we fail to find a
consistent solution. Both have multiple minima (not all of them
evident in this two dimensional contour plot). Because of the non-uniform time cadence
of HIP 20087 observations and a smaller number of measurements (26 vs. 32
for HIP 1366), the $\chi^2$ map for this star is more complex.  Another feature emerging from 
these plots is that the $\chi^2$
minimization favors high eccentricity solutions. This is not a feature
of this minimization algorithm, because any $\chi^2$-based global optimization
algorithm is biased toward high eccentricity in poorly conditioned problems. As noted by D. Pourbaix 
(2005)\footnote{http://wwwhip.obspm.fr/gaia/dms/texts/DMS-DP-02.pdf/},
the effective number of degrees of freedom goes up at high eccentricity, making it possible to find a fit with anomalously small
residuals if the number of observations is small.  Possible
solutions to this problem would be to introduce a penalty
function for high eccentricity solutions or use Bayesian approach with
a low prior probability of high eccentricity solutions. We
decided against using these techniques in this work to keep the algorithm as
simple as possible. Existing Hipparcos data do not preclude high
eccentricity solutions; and rigorous estimation of prior eccentricity
distributions is a project beyond the scope of this paper.

\section{Comparison with known binary solutions}
\label{known.sec}

\begin{figure}[htbp]
  \centering
  \includegraphics[angle=0,width=0.9\textwidth]{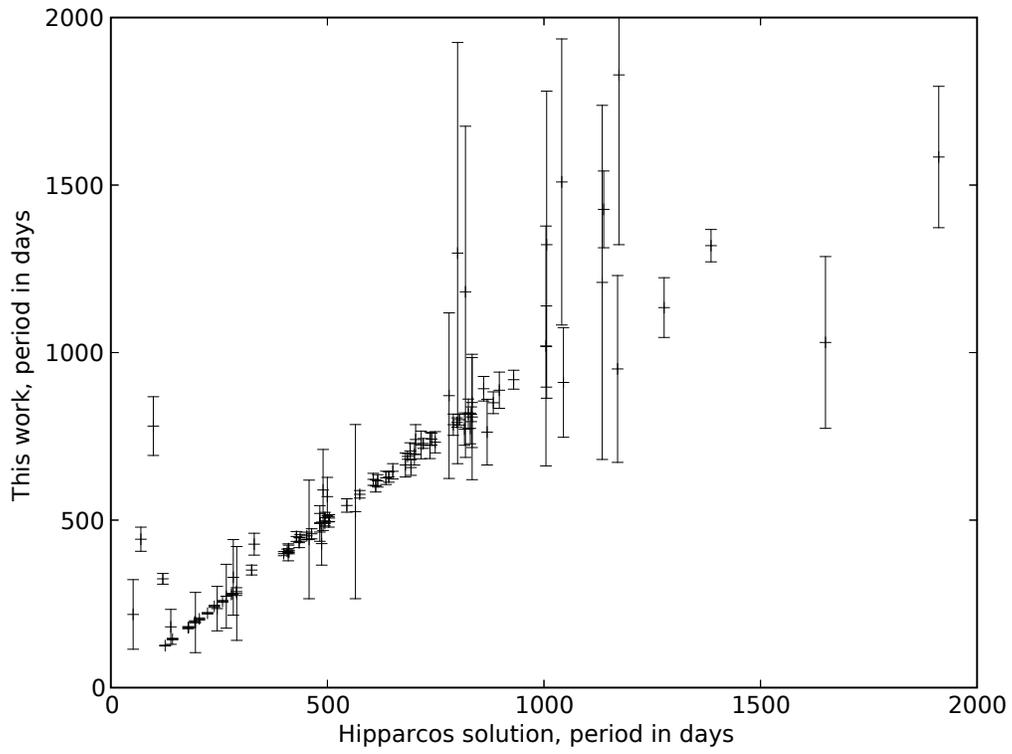}
  \caption{Comparison of orbital periods estimated by the genetic evolution algorithm versus periods in 
the Hipparcos catalog of orbital solutions. Only FAST consortium data are shown.}
  \label{fig:compare_periods}
\end{figure}

To verify the algorithm, we run it on a dataset of
235 stars with known orbital solutions in Hipparcos. The initial period, eccentricity
and periastron time for these Hipparcos solutions were often adopted from spectroscopic
data. In our work, we assume no prior knowledge of any orbit elements and compute the fits from scratch. 
The genetic algorithm is initialized with 30 points randomly
distributed in phase space and differential evolution simulations are
carried out for 100 generations for each star. The process
is repeated 1000 times for each star with input data modified by
adding normally distributed random numbers with standard deviations
equal to the formal errors. The results of this verification analysis are shown in
Fig. \ref{fig:compare_periods}, where the period of orbiting binaries derived by us and their
95\% confidence intervals are put to comparison with the periods from the Hipparcos catalog. 
The genetic algorithm fails to converge for four long period stars HIP 32349, 5336, 37279 and 95501.

The majority of orbital solutions are in good agreement as far as the crucial parameter $P$ is concerned. It is to
a large degree an external check for our method, since the Hipparcos orbital solutions are largely based on
more accurate spectroscopic orbits for the parameters in common (most notably, for $P$ and $e$). 
It is recalled that no apriori information
or constraints are used in our algorithm. With a few exceptions, the periods shorter than 1000 d appear to
be reliable. Binaries with longer periods exceeding the observation time span (typically, 3.2 yr) obtain uncertain
solutions. Such systems will benefit from using spectroscopic  or interferometric constraints. Four binaries with
short periods in Hipparcos ($P< 100$ d) obtain much longer periods in our unconstrained fits. These could be
hierarchical multiple systems, for which radial velocity monitoring tends to pick up the short-period, high velocity
amplitude signal, while the astrometric method is more sensitive to long-period signals of large angular excursions.

\section{Comparison with traditional optimization methods}
\label{known.opt.sec}

To compare the performance of our genetic optimization algorithm with
traditional optimization methods, we implement a brute force grid
search optimization algorithm and apply it to the same set of 235 stars with known orbital fits in
Hipparcos. The minimal value of $\chi^2$ is obtained by Powell optimization algorithm with initial starting points for
eccentricity at 0 and 0.5,  18 starting points for period spaced evenly in
logarithmic scale between 0.1 years and 30 years, and 4 starting points for
$T_0$ spaced evenly between 0 and $P$. This number of starting points $18\cdot 4 \cdot 2=144$ corresponds
to the lowest empirically determined density of the grid that still provides the highest possible rate of
correct solutions. The Powell minimization is iterated until the relative change in $\chi^2$ is less than 0.01.
The correct solutions are defined as solutions for which the fitted period is within a factor of 1.5 of
the spectroscopically derived period. For the FAST consortium data, the grid optimization routine yields 144
correct solutions, requiring on average 19500 function estimations. 

The differential evolution optimization algorithm with 30 initial points
and 30 generations produces 152 correct solutions, with 140 solutions
identical to the grid algorithm solutions. Each DE solution is followed by a Powell
optimization on the remaining linear parameters with the same 0.01 tolerance on $\chi^2$, totaling an average of 50 function
estimates.

In general, the DE global optimization algorithm requires approximately 20
times less function estimations to find the global optimum. Since the search space
is only 3-dimensional, the advantage of the DE algorithm is not overwhelming. This algorithm, however, allows us to use CPU-intensive
methods for estimating errors of free parameters, which is an important and subtle aspect of orbit
optimization. It requires days of CPU time instead of months on one of authors laptop. For problems with larger
number of dimensions, the DE algorithm may be even more attractive. It should be noted that the rather
low density of starting parameter grid in the grid optimization technique stems from the moderate intrinsic accuracy
of Hipparcos data (a few milliarcseconds), which is comparable to the detectable orbit size. The future space astrometry
missions (SIM and Gaia) will operate at a factor of $10^3$ higher accuracies, dramatically increasing the density of
the initial grid.

\section{Unconstrained solutions for stochastic stars}
\label{x.sec}
 
We further indiscriminately applied the genetic optimization algorithm to all Hipparcos stars with stochastic solutions.
Among the solutions demonstrating stable convergence, we selected 65 stars with confidence above 99\%, estimated as described
in section \ref{conf.sec}. A literature search for these 65 stars resulted in 11 systems with sufficiently accurate
spectroscopic orbits, listed in Table~\ref{known.tab}. Astrometrically estimated periods are usually quite close
to the more accurate spectroscopic periods at $P< 1000$ d. A few discrepant cases are present when
the periods from FAST and NDAC data are different by a significant fraction of the true quantity. Generally, the FAST
solutions seem to be more reliable. Probably, the FAST consortium used stricter criteria in treating statistical
outliers, occasional observations deviating far from the true astrometric abscissa, than the NDAC consortium.
Such outliers can disturb the nonlinear optimization process, generating
multiple wrinkles and gradients in the $\chi^2$ function. But some of the discrepant cases find more
interesting astrophysical explanation.

The star {\bf HIP 84949 = V819 Her} is a complex triple system. The inner binary is eclipsing of Algol type with
a period of 2.23 d. The outer companion of G8IV-III type is an active spotty star rotating with a period of
about 86 d \citep{vanh}. It can hardly be a coincidence that our best NDAC solution produced a period of 89 d.
In a binary system with a variable component, the photocenter moves along the line connecting the components
synchronously with the light curve, the effect known as Variability Imposed Movers (VIM).
The magnitude of the VIM astrometric excursion depends on the angular separation, magnitude difference and
the amplitude of variability. Accidentally, the VIM effect in HIP 84949 generated an orbital fit with the smallest
reduced $\chi^2$ with NDAC data, which was not at all a typical draw, since all the percentiles correspond
to long-period fits, i.e., the real outer orbit. The FAST solutions are not affected by the optical variability. This case
exemplifies the difficulties and hazards of purely astrometric orbital solutions for complicated multiple systems with
variable components.

Another piece of evidence that most of our discoveries are real long-period binaries is the high rate of occurrence
of accelerating stars \citep{maka}. This is not an independent check though, because apparent accelerations are
perceived either from the same Hipparcos observations, or from a comparison of Hipparcos and Tycho-2
proper motions. We find 19 stars in common with the two catalogs of accelerating stars in \citep{maka}.

Finally, the extensive Geneva-Copenhagen (GC) spectroscopic survey
\citep{nord}, and other radial velocity observations in the literature
provide a truly external verification of the binary nature of some of
our solutions.  The GC survey only includes Hipparcos stars of F, G
and K type. In some cases, only one RV measurement is available, so
that orbital motion is impossible to establish. In most cases, only a
few observations are available, which is not enough to derive a
orbital fit, but is sufficient to detect orbital motion. Combining
these data with Hipparcos astrometric data might be a subject for
future work.  Two indicators in the GC catalog are especially useful
in this respect: the probability of constant RV and a flag for
spectroscopic binaries.  Among the 54 new binaries from our analysis,
24 have these indicators of variable RV in the GC catalog.  A few more
stars have been known as spectroscopic binaries from earlier
investigations, but without reliable orbit estimations. Such is the case
with {\bf HIP 23221 = 63 Eri}, which is orbited by a white dwarf companion,
{\bf HIP 25732 = HD 3615} \citep{gre}, {\bf HIP 38146 = HD 63660}, a
G0III star \citep{demed}, {\bf HIP 43352 = HD 75605}, an UMa member
\citep{dem02}, {\bf HIP 62512 = HD 111456}, another UMa member
\citep{fre}, and {\bf HIP 76006 = HD 138525}, an F6 giant
\citep{dem02}.

\section{Overview of Astrometric Binaries}
\label{kin.sec}
The 54 previously unknown binary systems (Tables~\ref{ndac.tab.1}--\ref{fast.tab.1}) range in spectral type from A2 to M0.
Some of the early type stars stand out by their chemical peculiarities, most notably the $\lambda$-Boo type
star HIP 32607 = $\alpha$ Pic, the Am star HIP 30342 = $\nu$ Pic, and the Ap star HIP 1366 = HD 1280. 
\begin{itemize}
\item {\bf \object{HIP 1366} = HD 1280.} This star may also be quite young. Non-magnetic Ap stars are often found in SB2 systems,
and their companions are usually magnetic Am stars. This makes the subtle discrimination of Ap and Am
stars complicated for unresolved systems. \citet{car} derived spectroscopic orbits for 16 Ap stars, four of
which may be Am stars. They found orbital periods from 1.6--2422 d, which makes our solution, 1017 d,
a rather common case. Generally, \citet{car} find distributions of orbital elements similar to those of field
main-sequence stars, except for the distinct deficit of systems with $P<3$ d. The latter fact is explained
in terms of the tidal synchronization in tight binaries, which helps to retain high rotational velocities preventing
the occurrence of chemical peculiarity. From our solution based on FAST data, the visible semimajor axis
of the photocenter is about 1 AU, implying a rather massive companion, perhaps also an A star. HD 1280 has not been
identified as SB.

\item {\bf \object{HIP 30342} = HD 45229.} Also known as $\nu$ Pic, this Am and chemically peculiar star has drawn
considerable attention in the literature. Although Am stars are often found in SB2 systems \citep{abt}, ours is
the first indication of the binary nature of this object. Using our FAST parameters for $a$, $\pi$ and $P$, we
estimate the mass of the companion at only $\approx 1/4$ of the primary mass. The FAST corrected mass
function is $(M_2^3/M_{\rm tot}^2)_{\rm FAST}=0.09$.

\item {\bf \object{HIP 32607} = HD 50241.} Alias $\alpha$ Pic, a well-studied, bright $\lambda$ Boo-type star, not previously
known to be binary. Our derived $a$ is close to 1 AU, indicating a large mass ratio. This complies with the
suggestion of \citep{far04, far05} that many (may be all) $\lambda$ Boo stars are unresolved binaries with
composite spectra, which explain the observed chemical underabundances. The testable prediction here
is that the unresolved companions have similar brightness to the primaries, lest their contribution to the
observed spectrum is too small to account for the peculiarities of spectral lines. An alternative hypothesis of
the $\lambda$ Boo phenomenon is based on the gas-dust separation in stellar envelopes \citep[e.g.,][]{and}.
\end{itemize}

A few primaries in our sample are nearby late-type stars, whose companions can be low-mass dwarfs from the
bottom of the main sequence. A large number of M dwarfs and substellar objects (L and T type) are expected
to roam the solar vicinity, but relatively few have been identified. The nearest low-mass stars, because of their
intrinsic dimness, are traditionally important objects of investigation. Substellar companions are of particular
interest since the rate of such `failed stars' in spectroscopic binaries is currently considered to be too low
in comparison with the theoretical mass function expectation (the `brown dwarf desert' problem).

\begin{itemize}
\item {\bf HIP 38910.} Our solutions for $a$ and $P$ from FAST and NDAC are discrepant for this K5V star, probably because of
the very long period $>6$ yr. It seems that only a small segment of the orbit is present in the Hipparcos
data. The parallax is quite stable nevertheless, at 51-52 mas, that is 3-4 mas smaller than the value given in the
catalog. Our NDAC-based solution seems the more reasonable for this system, from which a small mass ratio is
estimated. The companion to HIP 38910 can be a nearby M dwarf.

\item {\bf HIP 84223 = GJ 1213.} A nearby K7 dwarf with apparently a long period. The updated parallaxes are above
40 mas, which warrants keeping this star in the Gliese catalog. With a period of several years and $a_a$ probably less
than 1 AU, the companion can be a late M dwarf. 

\item {\bf HIP 107143 = GJ 836.3.} This K3Vp dwarf has a stable solution with $a_a=33$ mas and $P\approx 1350$ d.
The updated parallax stays just above 40 mas. The companion can be a late K or early M dwarf.

\item {\bf HIP 118212 = GJ 913.} The latest star on our list, this nearby M0 dwarf stands out by its updated parallax of 68 mas
which is significantly larger than the cataloged value of 58 mas.  Our orbital fit is relatively
robust, suggesting a late M dwarf as a companion. 

\end{itemize}

Four early-type stars, HIP 12225 = HD 16555 = $\eta$ Hor (A6V), HIP 32607 = HD 50241 (A7IV), HIP 39903 = HD 68456 (F5V),
and HIP 62512 = GJ 9417 (F5V) are prominent X-ray sources detected by Rosat. Such objects are
occasionally found among the general field stars, as well as in nearby open clusters. Since the magnetic activity of the
corona is believed to be suppressed in such massive stars because of the vanishing convective zone, it
has been suggested that unresolved binary companions of smaller mass are the main contributors
in the detected X-ray flux.  Finding more astrometric binaries bolsters this surmise. The latter star, {\bf HIP 62512}
is of special interest. Beside the strong X-ray radiation, it is also an extreme UV source and a member of the $\approx
300$ Myr old UMa kinematic group. An astrometric excursion of roughly 1 AU coupled with a $\approx 4$ yr period
we determine for this system imply a mass ratio of 0.5. Thus, the companion can be a young, hot white dwarf.
Interestingly, we find three more UMa nucleus members in our sample, HIP 43352 = HD 75605, HIP 49546 and 
HIP 61100 = GJ 1160, the latter is a known spectroscopic binary (section \ref{known.sec}). With these new
addition to the list of UMa binaries, the binarity rate in the nucleus of this group becomes uncommonly high.

The star {\bf HIP 21965 = HD 30051} of spectral type F2/3 is suspected to be very young. Accurate age determination
in the upper half of the main sequence is difficult. Our orbital fit suggests a mass ratio of about 0.3, so that the
companion should have subsolar mass. If resolved, it can yield an accurate isochrone age for the system.

For {\bf HIP 23221 = HD 32008 = 63 Eri} we obtain a similar orbit to the previous system, with a slightly longer
period $P\approx 850$ d. This popular star has in fact been known as a spectroscopic binary, but a poor orbit
has been available in the literature. The secondary component is a white dwarf. Our orbital fit suggests a fairly
small mass ratio.

\section{Summary}

We present a method for unconstrained optimization of double stars
orbital parameters, based on an adaptation of the genetic evolution algorithm optimization and
Monte-Carlo simulations for error estimation. This method can
become a useful tool for estimating orbital parameters of binary stars and planet hosts for experiments
like SIM, which are expected to observe a limited number of pre-selected targets.
Experiments like Gaia are expected to observe much higher numbers of
objects and requirements for computing time are likely to become more
stringent. However, as the total processing time for the computations in this paper was
about two weeks of CPU time on a medium range laptop, we expect the entire
Gaia astrometric orbital estimation to become feasible on large scale
scientific clusters in the near future. This ``embarrassingly parallel''
algorithm is ideally suited for running in a grid environment and it will
take advantage of the ever increasing computer performance. We believe
that simplicity and ease of implementation of our algorithm outweigh the relatively high CPU time
requirements.

The 65 stars from the Hipparcos stochastic solution annex are good
candidates for follow-up spectroscopic studies, which are expected to yield more accurate orbital parameters
constraining the astrometric fits.

{\it Facility:} \facility{HIPPARCOS}

\appendix

\section{Linearized equations for orbital fits} \noindent
\label{A1.sec}
The Hipparcos Intermediate Astrometry Data (HIAD), the basic data
set to derive astrometric orbits, contains partial derivatives of
the star abscissa with respect the five astrometric parameters of the standard
model in equatorial coordinates,
\begin{eqnarray}
d_1 &=\partial a_i / \partial \alpha_* \\
d_2 &=\partial a_i / \partial \delta \\ 
d_3 &=\partial a_i / \partial \Pi \\
d_4 &=\partial a_i / \partial \mu_{\alpha *} \\ 
d_5 &=\partial a_i / \partial \mu_{\delta} 
\end{eqnarray}
where $a_i$ is the abscissa in $i$th observation of a given star, $\alpha$
and $\delta$ are the equatorial coordinates, $\alpha_*=\alpha\cos \delta$, $\Pi$ is the parallax, and
$\mu_{\alpha *}=\mu_{\alpha}\cos \delta$ and $ \mu_{\delta}$ are the
orthogonal proper motion components. The abscissa is defined as a
great circle arc connecting an arbitrary chosen reference zero-point
and the star on a fixed great circle (close to the scan circle).

In the small-angle approximation, a linearized equation for the
observed abscissa difference $\Delta a_i=a_{\rm obs} - a_{\rm calc}$,
can be written as
\eb
d_1\Delta x+d_2 \Delta y+d_3 \Delta \Pi+d_4 \Delta \mu_x +d_5 \Delta \mu_y
+ d_1 \sum_j\frac{\partial x}{\partial \epsilon_j}\Delta \epsilon_j
 + d_2 \sum_j\frac{\partial y}{\partial \epsilon_j}\Delta \epsilon_j
= \Delta a_i,
\label{des.eq}
\ee
where $\epsilon_j$ are the elements of the vector of seven orbital
elements, ${\bf \epsilon}= [a_a, P, e, T_0, \omega, \Omega, i]$.
Per usual, $a_a$ is the semimajor axis of the apparent orbital
motion (i.e., the motion of the photocenter), $P$ is the
orbital period in years, $e$ is the eccentricity, $T_0$ is the
periastron time, $\omega$ is the longitude of the ascending node
in the plane of orbit in radians, $\Omega$ is the position angle of the node in the plane
of projection, and $i$ is the inclination of the orbit (zero for face-on orbits).
In this equation, the notations $\alpha *$ and $\delta$ were changed to
$x$ and $y$, respectively, to make them consistent with the traditionally
used tangential coordinates for apparent orbits. Eq.~\ref{des.eq} holds
only in the vicinity of a certain point in the 12D parameter space
$\{\alpha, \delta, \Pi, \mu_{\alpha *},\mu_{\delta},{\bf \epsilon}\}$,
as long as the corrections to these parameters remain small.

The apparent motion of a binary in the plane of
celestial projection is described by \citep{hei}:
\begin{eqnarray}
x &=&A(\cos E-e)+F\sqrt{1-e^2}\sin E \\ \nonumber
y &=&B(\cos E-e)+G\sqrt{1-e^2}\sin E 
\label{coo.eq}
\end{eqnarray}
where $x$ and $y$ are the tangential coordinates,
and $E$ is the eccentric anomaly related to the mean anomaly $M$ by
Kepler's equation
\eb
M=2\pi\frac{T-T_0}{P}=E-e\sin E.
\ee
The Thiele-Innes constants are related to the remaining orbital
elements by
\begin{eqnarray}
A &=&a_a(\cos \omega\cos \Omega -\sin \omega \sin \Omega \cos i) \\ \nonumber
B &=&a_a(\cos \omega\sin \Omega +\sin \omega \cos \Omega \cos i) \\ \nonumber
F &=&a_a(-\sin \omega\cos \Omega -\cos \omega \sin \Omega \cos i) \\ \nonumber
G &=&a_a(-\sin \omega\sin \Omega +\cos \omega \cos \Omega \cos i), 
\end{eqnarray}
where $a_a$ is the apparent semimajor axis of the photocenter, $\omega$ is the periastron longitude,
$\Omega$ is the node and $i$ is the orbit inclination ($i=90\degr$ is an edge-on orbit).
 
\section{Covariances with linearized equations} \noindent
Explicitly, partial derivatives $\frac{\partial x}{\partial \epsilon_j}$ and $\frac{\partial y}{\partial \epsilon_j}$ in eq.~\ref{des.eq} are
\begin{eqnarray}
\frac{\partial x}{\partial a_a}&=&\frac{x}{a_a}\\ \nonumber
\frac{\partial y}{\partial a_a}&=&\frac{y}{a_a}\\ \nonumber
\frac{\partial x}{\partial P}&=&[-A\sin E+F\sqrt{1-e^2}\cos E]\frac{\partial E}{\partial P}\\ \nonumber
\frac{\partial y}{\partial P}&=&[-B\sin E+G\sqrt{1-e^2}\cos E]\frac{\partial E}{\partial P}\\ \nonumber
\frac{\partial x}{\partial e}&=&-A(1+\sin E\frac{\partial E}{\partial e})+F\frac{\sin E(\cos E-e)}{\sqrt{1-e^2}(1-e\cos E)}\\ \nonumber
\frac{\partial y}{\partial e}&=&-B(1+\sin E\frac{\partial E}{\partial e})+G\frac{\sin E(\cos E-e)}{\sqrt{1-e^2}(1-e\cos E)}\\ \nonumber
\frac{\partial x}{\partial T_0}&=&[-A\sin E+F\sqrt{1-e^2}\cos E]\frac{\partial E}{\partial T_0}\\ \nonumber
\frac{\partial y}{\partial T_0}&=&[-B\sin E+G\sqrt{1-e^2}\cos E]\frac{\partial E}{\partial T_0}\\ \nonumber
\frac{\partial x}{\partial \omega}&=&(\cos E-e)\frac{\partial A}{\partial \omega}+\sqrt{1-e^2}\sin E\frac{\partial F}{\partial \omega}\\ \nonumber
\frac{\partial y}{\partial \omega}&=&(\cos E-e)\frac{\partial B}{\partial \omega}+\sqrt{1-e^2}\sin E\frac{\partial G}{\partial \omega}\\ \nonumber
\frac{\partial x}{\partial \Omega}&=&(\cos E-e)\frac{\partial A}{\partial \Omega}+\sqrt{1-e^2}\sin E\frac{\partial F}{\partial \Omega}\\ \nonumber
\frac{\partial y}{\partial \Omega}&=&(\cos E-e)\frac{\partial B}{\partial \Omega}+\sqrt{1-e^2}\sin E\frac{\partial G}{\partial \Omega}\\ \nonumber
\frac{\partial x}{\partial i}&=&(\cos E-e)\frac{\partial A}{\partial i}+\sqrt{1-e^2}\sin E\frac{\partial F}{\partial i}\\ \nonumber
\frac{\partial y}{\partial i}&=&(\cos E-e)\frac{\partial B}{\partial i}+\sqrt{1-e^2}\sin E\frac{\partial G}{\partial i},
\label{par1.eq}
\end{eqnarray}
where
\begin{eqnarray}
\frac{\partial E}{\partial P}&=&\frac{2\pi (T-T_0)}{P^2(e\cos E-1)}\\ \nonumber
\frac{\partial E}{\partial e}&=&\frac{\sin E}{1-e\cos E}\\ \nonumber
\frac{\partial E}{\partial T_0}&=&\frac{-2\pi}{P(1-e\cos E)}\\ \nonumber
\frac{\partial A}{\partial \omega}&=&F\\ \nonumber
\frac{\partial B}{\partial \omega}&=&G\\ \nonumber
\frac{\partial F}{\partial \omega}&=&-A\\ \nonumber
\frac{\partial G}{\partial \omega}&=&-B\\ \nonumber
\frac{\partial A}{\partial \Omega}&=&-B\\ \nonumber
\frac{\partial B}{\partial \Omega}&=&A\\ \nonumber
\frac{\partial F}{\partial \Omega}&=&-G\\ \nonumber
\frac{\partial G}{\partial \Omega}&=&F\\ \nonumber
\frac{\partial A}{\partial i}&=&a_a\sin\omega \sin\Omega \sin i\\ \nonumber
\frac{\partial B}{\partial \omega}&=&-a_a\sin\omega \cos\Omega \sin i\\ \nonumber
\frac{\partial F}{\partial \omega}&=&a_a\cos\omega \sin\Omega \sin i\\ \nonumber
\frac{\partial G}{\partial \omega}&=&-a_a\cos\omega \cos\Omega \sin i  \vspace{5mm}.
\label{par2.eq}
\end{eqnarray}

Using these relations, the design matrix $D$ can be computed by eq.~\ref{des.eq}, and the covariance
matrix $C=(D^T D)^{-1}$ is readily computed. Astrometric abscissae $a_i$ for the same star may
have different standard errors in HIAD, making it necessary to employ a weighted least-squares routine.
 
\acknowledgments
The research described in this paper was in part carried out at the Jet Propulsion 
Laboratory, California Institute of Technology, under a contract with the National 
Aeronautics and Space Administration. This research has made use of the SIMBAD database,
operated at CDS, Strasbourg, France.

\begin{deluxetable}{ r | r |  r |  r |  r |  r | r |  r |  r}
\tabletypesize{\scriptsize}
\tablecaption{Orbital solutions  from NDAC data. \label{ndac.tab.1}}
\tablewidth{0pt}
\startdata

\multicolumn{1}{c}{ HIP}& \multicolumn{1}{c}{$a_0$}  &  \multicolumn{1}{c}{$P$}  &  \multicolumn{1}{c}{$\epsilon$}  &  \multicolumn{1}{c}{$T_0$}  &  \multicolumn{1}{c}{$\Omega$}  &  \multicolumn{1}{c}{$\omega$}  &  \multicolumn{1}{c}{$i$}  &  \multicolumn{1}{c}{parallax}  \\
\multicolumn{1}{c}{}& \multicolumn{1}{c}{mas}  &  \multicolumn{1}{c}{d}  &  \multicolumn{1}{c}{}  &  \multicolumn{1}{c}{d}  &  \multicolumn{1}{c}{$\degr$}  &  \multicolumn{1}{c}{$\degr$}  &  \multicolumn{1}{c}{$\degr$}  &  \multicolumn{1}{c}{mas}  \\
\hline
 & &  &  &  &  &  &  &  \\
1366   & $25^{+32}_{-13}$   & $1033^{+91}_{-77}$   & $0.95^{+0.04}_{-0.40}$   & $238^{+343}_{-143}$   & $263^{+50}_{-156}$   & $89^{+175}_{-29}$   & $69^{+8}_{-22}$   & $14^{+1}_{-1}$   \\ 
3750   & $18^{+10}_{-3}$   & $1646^{+609}_{-321}$   & $0.74^{+0.25}_{-0.18}$   & $704^{+670}_{-569}$   & $80^{+6}_{-9}$   & $35^{+33}_{-14}$   & $77^{+6}_{-10}$   & $12^{+1}_{-1}$   \\ 
3865   & $10^{+5}_{-1}$   & $1142^{+146}_{-144}$   & $0.47^{+0.37}_{-0.25}$   & $861^{+251}_{-455}$   & $86^{+206}_{-48}$   & $82^{+76}_{-51}$   & $44^{+18}_{-13}$   & $11^{+1}_{-1}$   \\ 
6273   & $46^{+57}_{-10}$   & $2656^{+1893}_{-805}$   & $0.86^{+0.12}_{-0.18}$   & $1364^{+1855}_{-753}$   & $195^{+59}_{-117}$   & $105^{+167}_{-67}$   & $137^{+14}_{-22}$   & $34^{+2}_{-1}$   \\ 
6542   & $23^{+8}_{-3}$   & $1760^{+550}_{-280}$   & $0.65^{+0.20}_{-0.22}$   & $435^{+520}_{-198}$   & $138^{+13}_{-13}$   & $109^{+28}_{-19}$   & $54^{+8}_{-7}$   & $18^{+1}_{-1}$   \\ 
8525   & $19^{+21}_{-5}$   & $704^{+31}_{-25}$   & $0.90^{+0.09}_{-0.37}$   & $529^{+88}_{-94}$   & $233^{+37}_{-183}$   & $122^{+159}_{-74}$   & $72^{+11}_{-13}$   & $23^{+1}_{-1}$   \\ 
11925   & $16^{+19}_{-3}$   & $920^{+236}_{-38}$   & $0.71^{+0.23}_{-0.18}$   & $785^{+100}_{-623}$   & $344^{+14}_{-341}$   & $102^{+162}_{-17}$   & $94^{+6}_{-5}$   & $15^{+1}_{-1}$   \\ 
12062   & $34^{+39}_{-22}$   & $979^{+254}_{-209}$   & $0.97^{+0.02}_{-0.54}$   & $505^{+234}_{-343}$   & $203^{+38}_{-152}$   & $90^{+176}_{-22}$   & $78^{+6}_{-18}$   & $18^{+2}_{-2}$   \\ 
12225   & $24^{+3}_{-1}$   & $1148^{+369}_{-116}$   & $0.26^{+0.25}_{-0.17}$   & $939^{+186}_{-436}$   & $62^{+176}_{-5}$   & $88^{+56}_{-36}$   & $115^{+4}_{-4}$   & $22^{+1}_{-1}$   \\ 
12894   & $11^{+3}_{-1}$   & $861^{+27}_{-26}$   & $0.46^{+0.27}_{-0.18}$   & $547^{+139}_{-107}$   & $284^{+18}_{-170}$   & $106^{+166}_{-32}$   & $123^{+9}_{-9}$   & $16^{+1}_{-1}$   \\ 
17022   & $14^{+32}_{-3}$   & $1091^{+114}_{-74}$   & $0.68^{+0.31}_{-0.24}$   & $336^{+280}_{-172}$   & $112^{+163}_{-40}$   & $121^{+155}_{-80}$   & $69^{+12}_{-9}$   & $21^{+2}_{-2}$   \\ 
19832   & $21^{+4}_{-2}$   & $669^{+26}_{-25}$   & $0.31^{+0.25}_{-0.17}$   & $264^{+191}_{-109}$   & $288^{+62}_{-137}$   & $66^{+147}_{-40}$   & $56^{+11}_{-11}$   & $46^{+3}_{-3}$   \\ 
21386   & $20^{+2}_{-1}$   & $1031^{+33}_{-26}$   & $0.23^{+0.14}_{-0.12}$   & $858^{+112}_{-137}$   & $30^{+175}_{-9}$   & $89^{+97}_{-68}$   & $98^{+2}_{-2}$   & $26^{+1}_{-1}$   \\ 
21965   & $11^{+30}_{-2}$   & $710^{+40}_{-35}$   & $0.56^{+0.43}_{-0.31}$   & $515^{+106}_{-148}$   & $155^{+41}_{-54}$   & $100^{+53}_{-28}$   & $126^{+14}_{-21}$   & $16^{+1}_{-1}$   \\ 
23221   & $12^{+21}_{-4}$   & $829^{+81}_{-56}$   & $0.87^{+0.12}_{-0.45}$   & $273^{+383}_{-155}$   & $61^{+174}_{-23}$   & $97^{+83}_{-58}$   & $97^{+11}_{-9}$   & $18^{+1}_{-1}$   \\ 
23776   & $17^{+29}_{-8}$   & $709^{+110}_{-45}$   & $0.92^{+0.07}_{-0.31}$   & $190^{+150}_{-107}$   & $100^{+159}_{-12}$   & $91^{+22}_{-15}$   & $82^{+6}_{-10}$   & $28^{+1}_{-1}$   \\ 
25732   & $32^{+13}_{-6}$   & $2210^{+2014}_{-590}$   & $0.47^{+0.23}_{-0.23}$   & $845^{+1892}_{-496}$   & $279^{+9}_{-176}$   & $92^{+101}_{-71}$   & $61^{+6}_{-6}$   & $12^{+1}_{-1}$   \\ 
25918   & $51^{+14}_{-6}$   & $2723^{+2584}_{-963}$   & $0.59^{+0.17}_{-0.13}$   & $1410^{+2035}_{-949}$   & $117^{+173}_{-8}$   & $124^{+48}_{-101}$   & $53^{+9}_{-6}$   & $30^{+1}_{-1}$   \\ 
30223   & $17^{+30}_{-9}$   & $799^{+70}_{-73}$   & $0.92^{+0.07}_{-0.65}$   & $258^{+280}_{-147}$   & $137^{+187}_{-60}$   & $89^{+64}_{-30}$   & $66^{+13}_{-22}$   & $15^{+2}_{-2}$   \\ 
30342   & $10^{+33}_{-3}$   & $439^{+27}_{-23}$   & $0.75^{+0.24}_{-0.40}$   & $328^{+85}_{-306}$   & $163^{+82}_{-116}$   & $94^{+71}_{-15}$   & $107^{+18}_{-11}$   & $20^{+1}_{-1}$   \\ 
31703   & $41^{+34}_{-16}$   & $2901^{+3399}_{-1412}$   & $0.84^{+0.08}_{-0.11}$   & $1588^{+3392}_{-1114}$   & $99^{+16}_{-21}$   & $101^{+13}_{-10}$   & $132^{+10}_{-9}$   & $24^{+1}_{-1}$   \\ 
32607   & $36^{+15}_{-2}$   & $1618^{+1407}_{-325}$   & $0.39^{+0.35}_{-0.17}$   & $953^{+707}_{-640}$   & $24^{+5}_{-5}$   & $92^{+22}_{-44}$   & $118^{+3}_{-3}$   & $34^{+1}_{-1}$   \\ 
36399   & $22^{+74}_{-5}$   & $995^{+50}_{-39}$   & $0.77^{+0.22}_{-0.16}$   & $815^{+143}_{-767}$   & $253^{+26}_{-165}$   & $125^{+162}_{-31}$   & $130^{+13}_{-28}$   & $39^{+1}_{-1}$   \\ 
37606   & $21^{+30}_{-11}$   & $380^{+10}_{-11}$   & $0.93^{+0.06}_{-0.34}$   & $81^{+71}_{-45}$   & $202^{+16}_{-176}$   & $84^{+167}_{-26}$   & $99^{+18}_{-8}$   & $44^{+2}_{-2}$   \\ 
38018   & $19^{+3}_{-2}$   & $543^{+11}_{-11}$   & $0.60^{+0.16}_{-0.12}$   & $452^{+48}_{-40}$   & $250^{+16}_{-168}$   & $51^{+159}_{-28}$   & $46^{+8}_{-9}$   & $32^{+1}_{-1}$   \\ 
38146   & $13^{+3}_{-2}$   & $902^{+18}_{-20}$   & $0.50^{+0.28}_{-0.19}$   & $581^{+56}_{-105}$   & $187^{+158}_{-32}$   & $123^{+69}_{-109}$   & $42^{+11}_{-12}$   & $10^{+1}_{-1}$   \\ 
38596   & $25^{+23}_{-5}$   & $2427^{+2056}_{-758}$   & $0.66^{+0.26}_{-0.22}$   & $1250^{+1388}_{-889}$   & $319^{+18}_{-169}$   & $92^{+146}_{-48}$   & $66^{+10}_{-9}$   & $27^{+2}_{-2}$   \\ 
38910   & $39^{+9}_{-3}$   & $2041^{+1362}_{-398}$   & $0.31^{+0.29}_{-0.18}$   & $1130^{+797}_{-693}$   & $199^{+17}_{-166}$   & $121^{+89}_{-72}$   & $56^{+7}_{-9}$   & $52^{+1}_{-1}$   \\ 
38980   & $18^{+33}_{-6}$   & $894^{+196}_{-68}$   & $0.86^{+0.13}_{-0.44}$   & $458^{+207}_{-220}$   & $292^{+13}_{-174}$   & $87^{+171}_{-19}$   & $80^{+6}_{-8}$   & $36^{+1}_{-1}$   \\ 
39681   & $23^{+12}_{-3}$   & $1964^{+1730}_{-484}$   & $0.58^{+0.28}_{-0.37}$   & $945^{+1220}_{-678}$   & $59^{+12}_{-9}$   & $99^{+29}_{-30}$   & $62^{+10}_{-12}$   & $16^{+1}_{-2}$   \\ 
39903   & $26^{+1}_{-1}$   & $901^{+16}_{-15}$   & $0.07^{+0.06}_{-0.04}$   & $224^{+498}_{-143}$   & $175^{+172}_{-162}$   & $80^{+102}_{-56}$   & $23^{+6}_{-7}$   & $49^{+1}_{-1}$   \\ 
40015   & $14^{+19}_{-8}$   & $247^{+6}_{-9}$   & $0.97^{+0.02}_{-0.56}$   & $178^{+59}_{-163}$   & $107^{+184}_{-68}$   & $94^{+74}_{-18}$   & $108^{+22}_{-11}$   & $33^{+1}_{-1}$   \\ 
42916   & $23^{+4}_{-2}$   & $817^{+13}_{-17}$   & $0.68^{+0.12}_{-0.08}$   & $338^{+49}_{-61}$   & $154^{+66}_{-96}$   & $121^{+58}_{-70}$   & $148^{+10}_{-11}$   & $37^{+1}_{-1}$   \\ 
43352   & $12^{+37}_{-3}$   & $1180^{+253}_{-143}$   & $0.74^{+0.25}_{-0.20}$   & $423^{+515}_{-283}$   & $195^{+14}_{-172}$   & $93^{+176}_{-11}$   & $119^{+12}_{-21}$   & $14^{+1}_{-1}$   \\ 
49546   & $12^{+21}_{-3}$   & $584^{+81}_{-26}$   & $0.73^{+0.25}_{-0.32}$   & $436^{+107}_{-360}$   & $326^{+9}_{-176}$   & $82^{+170}_{-22}$   & $92^{+6}_{-5}$   & $15^{+1}_{-1}$   \\ 
50180   & $35^{+19}_{-4}$   & $1764^{+1496}_{-327}$   & $0.32^{+0.26}_{-0.18}$   & $978^{+789}_{-643}$   & $192^{+6}_{-178}$   & $150^{+132}_{-90}$   & $79^{+4}_{-6}$   & $29^{+1}_{-1}$   \\ 
54424   & $10^{+2}_{-1}$   & $1177^{+172}_{-165}$   & $0.50^{+0.19}_{-0.17}$   & $654^{+338}_{-296}$   & $278^{+42}_{-169}$   & $95^{+149}_{-42}$   & $135^{+12}_{-11}$   & $13^{+1}_{-1}$   \\ 
54746   & $14^{+23}_{-3}$   & $975^{+58}_{-44}$   & $0.88^{+0.11}_{-0.18}$   & $129^{+198}_{-86}$   & $79^{+24}_{-13}$   & $124^{+33}_{-25}$   & $68^{+10}_{-10}$   & $19^{+1}_{-1}$   \\ 
61100   & $35^{+26}_{-5}$   & $1270^{+145}_{-92}$   & $0.65^{+0.24}_{-0.14}$   & $137^{+917}_{-96}$   & $175^{+7}_{-10}$   & $76^{+14}_{-15}$   & $63^{+8}_{-6}$   & $41^{+1}_{-1}$   \\ 
62512   & $44^{+5}_{-2}$   & $1510^{+456}_{-133}$   & $0.16^{+0.25}_{-0.09}$   & $244^{+638}_{-143}$   & $31^{+177}_{-3}$   & $154^{+24}_{-126}$   & $106^{+2}_{-2}$   & $40^{+1}_{-1}$   \\ 
73440   & $13^{+22}_{-7}$   & $436^{+14}_{-18}$   & $0.94^{+0.05}_{-0.48}$   & $85^{+299}_{-60}$   & $229^{+43}_{-144}$   & $88^{+147}_{-37}$   & $70^{+13}_{-17}$   & $30^{+1}_{-1}$   \\ 
75401   & $22^{+12}_{-4}$   & $1171^{+630}_{-233}$   & $0.62^{+0.24}_{-0.21}$   & $753^{+410}_{-503}$   & $283^{+14}_{-170}$   & $101^{+127}_{-56}$   & $68^{+8}_{-9}$   & $19^{+1}_{-1}$   \\ 
76006   & $10^{+24}_{-4}$   & $584^{+27}_{-31}$   & $0.86^{+0.13}_{-0.36}$   & $342^{+97}_{-99}$   & $95^{+54}_{-25}$   & $97^{+43}_{-12}$   & $64^{+16}_{-19}$   & $15^{+1}_{-1}$   \\ 
78970   & $12^{+3}_{-1}$   & $868^{+54}_{-54}$   & $0.28^{+0.28}_{-0.15}$   & $398^{+265}_{-129}$   & $197^{+13}_{-173}$   & $91^{+149}_{-40}$   & $67^{+7}_{-7}$   & $22^{+1}_{-1}$   \\ 
80884   & $14^{+2}_{-1}$   & $992^{+93}_{-41}$   & $0.49^{+0.19}_{-0.16}$   & $321^{+620}_{-288}$   & $292^{+8}_{-176}$   & $109^{+167}_{-21}$   & $70^{+6}_{-7}$   & $12^{+1}_{-1}$   \\ 
84062   & $13^{+28}_{-2}$   & $746^{+41}_{-42}$   & $0.68^{+0.30}_{-0.31}$   & $108^{+568}_{-82}$   & $170^{+108}_{-151}$   & $94^{+155}_{-78}$   & $53^{+17}_{-13}$   & $35^{+1}_{-1}$   \\ 
84223   & $41^{+87}_{-12}$   & $4219^{+5366}_{-1992}$   & $0.76^{+0.23}_{-0.35}$   & $2705^{+5357}_{-1840}$   & $169^{+184}_{-165}$   & $105^{+59}_{-22}$   & $102^{+8}_{-7}$   & $46^{+2}_{-2}$   \\ 
84949   & $28^{+37}_{-10}$   & $2079^{+1136}_{-645}$   & $0.91^{+0.08}_{-0.16}$   & $808^{+1006}_{-535}$   & $127^{+16}_{-17}$   & $75^{+11}_{-16}$   & $71^{+10}_{-8}$   & $15^{+1}_{-1}$   \\ 
85852   & $9^{+2}_{-1}$   & $944^{+69}_{-51}$   & $0.36^{+0.30}_{-0.19}$   & $653^{+197}_{-375}$   & $211^{+134}_{-67}$   & $92^{+134}_{-60}$   & $125^{+12}_{-10}$   & $11^{+1}_{-1}$   \\ 
88848   & $40^{+19}_{-6}$   & $1948^{+1055}_{-436}$   & $0.67^{+0.19}_{-0.21}$   & $622^{+858}_{-476}$   & $250^{+5}_{-178}$   & $105^{+171}_{-16}$   & $79^{+3}_{-3}$   & $30^{+1}_{-1}$   \\ 
90355   & $13^{+24}_{-2}$   & $297^{+8}_{-9}$   & $0.51^{+0.46}_{-0.23}$   & $161^{+54}_{-40}$   & $177^{+46}_{-127}$   & $97^{+139}_{-54}$   & $134^{+14}_{-22}$   & $30^{+2}_{-2}$   \\ 
93137   & $13^{+6}_{-2}$   & $847^{+66}_{-41}$   & $0.59^{+0.29}_{-0.19}$   & $459^{+177}_{-105}$   & $33^{+8}_{-8}$   & $85^{+23}_{-21}$   & $80^{+6}_{-7}$   & $17^{+1}_{-1}$   \\ 
94347   & $14^{+4}_{-2}$   & $633^{+20}_{-20}$   & $0.45^{+0.23}_{-0.16}$   & $377^{+101}_{-72}$   & $35^{+120}_{-21}$   & $76^{+30}_{-26}$   & $143^{+11}_{-12}$   & $22^{+1}_{-1}$   \\ 
94802   & $18^{+1}_{-1}$   & $1069^{+47}_{-35}$   & $0.15^{+0.10}_{-0.08}$   & $571^{+163}_{-206}$   & $87^{+10}_{-8}$   & $77^{+65}_{-43}$   & $132^{+5}_{-5}$   & $23^{+1}_{-1}$   \\ 
97063   & $19^{+5}_{-2}$   & $1422^{+656}_{-177}$   & $0.36^{+0.28}_{-0.21}$   & $673^{+612}_{-356}$   & $201^{+92}_{-139}$   & $99^{+147}_{-60}$   & $36^{+11}_{-11}$   & $28^{+1}_{-1}$   \\ 
97690   & $9^{+5}_{-1}$   & $1007^{+108}_{-122}$   & $0.44^{+0.42}_{-0.25}$   & $544^{+300}_{-340}$   & $41^{+167}_{-19}$   & $87^{+85}_{-57}$   & $63^{+11}_{-12}$   & $11^{+1}_{-1}$   \\ 
98375   & $12^{+22}_{-5}$   & $285^{+14}_{-15}$   & $0.96^{+0.03}_{-0.47}$   & $135^{+101}_{-94}$   & $181^{+27}_{-162}$   & $97^{+161}_{-29}$   & $92^{+10}_{-9}$   & $21^{+1}_{-1}$   \\ 
103287   & $31^{+15}_{-3}$   & $1484^{+602}_{-258}$   & $0.50^{+0.34}_{-0.29}$   & $878^{+478}_{-690}$   & $339^{+5}_{-178}$   & $119^{+165}_{-28}$   & $86^{+5}_{-7}$   & $24^{+2}_{-2}$   \\ 
104440   & $68^{+82}_{-27}$   & $2145^{+3036}_{-914}$   & $0.70^{+0.10}_{-0.08}$   & $1073^{+2310}_{-716}$   & $13^{+180}_{-4}$   & $148^{+32}_{-135}$   & $90^{+2}_{-2}$   & $51^{+1}_{-1}$   \\ 
105969   & $13^{+2}_{-1}$   & $885^{+34}_{-36}$   & $0.34^{+0.26}_{-0.18}$   & $415^{+179}_{-145}$   & $109^{+178}_{-24}$   & $130^{+47}_{-107}$   & $133^{+11}_{-9}$   & $15^{+1}_{-1}$   \\ 
107143   & $36^{+9}_{-3}$   & $1371^{+562}_{-111}$   & $0.28^{+0.34}_{-0.17}$   & $699^{+653}_{-628}$   & $111^{+175}_{-30}$   & $77^{+95}_{-51}$   & $46^{+10}_{-9}$   & $41^{+2}_{-2}$   \\ 
110291   & $17^{+26}_{-8}$   & $757^{+82}_{-58}$   & $0.92^{+0.07}_{-0.34}$   & $420^{+295}_{-337}$   & $72^{+171}_{-27}$   & $85^{+86}_{-24}$   & $99^{+12}_{-8}$   & $18^{+3}_{-2}$   \\ 
113638   & $23^{+34}_{-15}$   & $356^{+11}_{-18}$   & $0.91^{+0.08}_{-0.53}$   & $206^{+113}_{-70}$   & $158^{+35}_{-51}$   & $88^{+32}_{-14}$   & $76^{+11}_{-24}$   & $30^{+3}_{-5}$   \\ 
117622   & $14^{+46}_{-2}$   & $1011^{+104}_{-65}$   & $0.59^{+0.40}_{-0.35}$   & $285^{+549}_{-193}$   & $28^{+26}_{-19}$   & $99^{+44}_{-31}$   & $126^{+15}_{-25}$   & $20^{+1}_{-2}$   \\ 
118212   & $35^{+4}_{-3}$   & $885^{+28}_{-17}$   & $0.56^{+0.12}_{-0.10}$   & $553^{+76}_{-63}$   & $110^{+6}_{-6}$   & $110^{+15}_{-17}$   & $115^{+5}_{-5}$   & $67^{+3}_{-3}$   \\ 
\enddata
\end{deluxetable}

\begin{deluxetable}{ r | r |  r |  r |  r |  r |  r | r |  r}
\tabletypesize{\scriptsize}
\tablecaption{Orbital solutions from FAST data. \label{fast.tab.1}}
\tablewidth{0pt}
\startdata
\multicolumn{1}{c}{HIP}& \multicolumn{1}{c}{$a_0$}  &  \multicolumn{1}{c}{$P$}  &  \multicolumn{1}{c}{$\epsilon$}  &  \multicolumn{1}{c}{$T_0$}  &  \multicolumn{1}{c}{$\Omega$}  &  \multicolumn{1}{c}{$\omega$}  &  \multicolumn{1}{c}{$i$}  &  \multicolumn{1}{c}{parallax}  \\
\multicolumn{1}{c}{}& \multicolumn{1}{c}{mas}  &  \multicolumn{1}{c}{d}  &  \multicolumn{1}{c}{}  &  \multicolumn{1}{c}{d}  &  \multicolumn{1}{c}{$\degr$} &  \multicolumn{1}{c}{$\degr$}  &  \multicolumn{1}{c}{$\degr$}  &  \multicolumn{1}{c}{mas}  \\
\hline
 & &  &  &  &  &  &  &  \\
1366   & $12^{+9}_{-2}$   & $1006^{+88}_{-102}$   & $0.62^{+0.31}_{-0.25}$   & $272^{+480}_{-174}$   & $294^{+18}_{-171}$   & $70^{+149}_{-42}$   & $63^{+10}_{-10}$   & $15^{+1}_{-1}$   \\ 
3750   & $17^{+10}_{-3}$   & $1640^{+809}_{-403}$   & $0.82^{+0.17}_{-0.20}$   & $779^{+615}_{-575}$   & $82^{+7}_{-6}$   & $36^{+36}_{-21}$   & $86^{+5}_{-8}$   & $13^{+1}_{-1}$   \\ 
3865   & $24^{+20}_{-13}$   & $1325^{+371}_{-277}$   & $0.97^{+0.02}_{-0.28}$   & $527^{+599}_{-432}$   & $54^{+236}_{-42}$   & $82^{+10}_{-38}$   & $68^{+9}_{-24}$   & $10^{+1}_{-1}$   \\ 
6273   & $55^{+67}_{-15}$   & $3261^{+3108}_{-1316}$   & $0.84^{+0.15}_{-0.28}$   & $1995^{+3090}_{-1236}$   & $135^{+131}_{-37}$   & $95^{+85}_{-78}$   & $128^{+14}_{-15}$   & $34^{+2}_{-2}$   \\ 
6542   & $22^{+8}_{-3}$   & $1807^{+570}_{-282}$   & $0.62^{+0.22}_{-0.26}$   & $441^{+555}_{-224}$   & $144^{+10}_{-11}$   & $99^{+23}_{-16}$   & $50^{+7}_{-7}$   & $16^{+1}_{-1}$   \\ 
8525   & $19^{+27}_{-6}$   & $731^{+36}_{-28}$   & $0.86^{+0.13}_{-0.45}$   & $576^{+84}_{-130}$   & $231^{+30}_{-191}$   & $105^{+130}_{-70}$   & $72^{+8}_{-11}$   & $24^{+2}_{-1}$   \\ 
11925   & $13^{+7}_{-2}$   & $974^{+181}_{-61}$   & $0.64^{+0.25}_{-0.20}$   & $252^{+626}_{-199}$   & $350^{+8}_{-347}$   & $125^{+161}_{-20}$   & $97^{+6}_{-5}$   & $16^{+1}_{-1}$   \\ 
12062   & $16^{+50}_{-4}$   & $1028^{+123}_{-166}$   & $0.78^{+0.21}_{-0.28}$   & $348^{+335}_{-200}$   & $184^{+64}_{-113}$   & $90^{+118}_{-37}$   & $51^{+27}_{-17}$   & $19^{+1}_{-1}$   \\ 
12225   & $25^{+5}_{-2}$   & $1247^{+1148}_{-175}$   & $0.31^{+0.32}_{-0.21}$   & $976^{+421}_{-707}$   & $53^{+174}_{-8}$   & $71^{+100}_{-34}$   & $121^{+5}_{-4}$   & $21^{+1}_{-1}$   \\ 
12894   & $10^{+3}_{-1}$   & $862^{+33}_{-31}$   & $0.35^{+0.36}_{-0.19}$   & $513^{+238}_{-135}$   & $281^{+23}_{-166}$   & $85^{+147}_{-49}$   & $119^{+8}_{-8}$   & $15^{+1}_{-1}$   \\ 
17022   & $15^{+8}_{-2}$   & $1119^{+98}_{-63}$   & $0.63^{+0.28}_{-0.17}$   & $423^{+234}_{-169}$   & $83^{+184}_{-13}$   & $96^{+81}_{-82}$   & $65^{+7}_{-6}$   & $20^{+1}_{-1}$   \\ 
19832   & $21^{+7}_{-3}$   & $662^{+34}_{-34}$   & $0.32^{+0.38}_{-0.18}$   & $383^{+231}_{-334}$   & $170^{+37}_{-136}$   & $75^{+95}_{-32}$   & $47^{+13}_{-11}$   & $43^{+3}_{-4}$   \\ 
21386   & $20^{+1}_{-1}$   & $1044^{+29}_{-23}$   & $0.23^{+0.13}_{-0.11}$   & $909^{+89}_{-134}$   & $203^{+5}_{-178}$   & $59^{+136}_{-38}$   & $100^{+2}_{-2}$   & $25^{+1}_{-1}$   \\ 
21965   & $10^{+2}_{-1}$   & $718^{+21}_{-20}$   & $0.30^{+0.27}_{-0.17}$   & $456^{+117}_{-331}$   & $167^{+167}_{-13}$   & $104^{+48}_{-41}$   & $123^{+8}_{-7}$   & $15^{+1}_{-1}$   \\ 
23221   & $17^{+22}_{-7}$   & $862^{+71}_{-53}$   & $0.93^{+0.06}_{-0.22}$   & $257^{+123}_{-102}$   & $40^{+11}_{-9}$   & $109^{+27}_{-12}$   & $93^{+7}_{-5}$   & $19^{+1}_{-1}$   \\ 
23776   & $12^{+26}_{-4}$   & $718^{+35}_{-38}$   & $0.84^{+0.15}_{-0.34}$   & $245^{+99}_{-86}$   & $101^{+156}_{-11}$   & $96^{+34}_{-23}$   & $77^{+9}_{-12}$   & $28^{+1}_{-1}$   \\ 
25732   & $31^{+19}_{-8}$   & $2707^{+2803}_{-1079}$   & $0.50^{+0.24}_{-0.30}$   & $1341^{+2667}_{-964}$   & $113^{+170}_{-10}$   & $146^{+31}_{-128}$   & $62^{+10}_{-7}$   & $12^{+1}_{-1}$   \\ 
25918   & $60^{+24}_{-10}$   & $3109^{+2525}_{-1086}$   & $0.53^{+0.19}_{-0.13}$   & $1560^{+2098}_{-1084}$   & $123^{+19}_{-12}$   & $124^{+42}_{-34}$   & $52^{+9}_{-6}$   & $31^{+1}_{-1}$   \\ 
30223   & $10^{+30}_{-2}$   & $796^{+78}_{-64}$   & $0.56^{+0.43}_{-0.32}$   & $240^{+261}_{-126}$   & $146^{+170}_{-42}$   & $88^{+78}_{-56}$   & $51^{+22}_{-13}$   & $14^{+2}_{-2}$   \\ 
30342   & $6^{+20}_{-1}$   & $442^{+19}_{-25}$   & $0.46^{+0.53}_{-0.27}$   & $206^{+162}_{-171}$   & $112^{+129}_{-72}$   & $93^{+75}_{-38}$   & $119^{+17}_{-17}$   & $20^{+1}_{-1}$   \\ 
31703   & $36^{+34}_{-13}$   & $2676^{+3382}_{-1233}$   & $0.83^{+0.09}_{-0.11}$   & $1382^{+3371}_{-981}$   & $102^{+27}_{-50}$   & $106^{+25}_{-26}$   & $139^{+11}_{-10}$   & $25^{+1}_{-1}$   \\ 
32607   & $35^{+11}_{-2}$   & $1528^{+1013}_{-264}$   & $0.37^{+0.33}_{-0.16}$   & $915^{+615}_{-613}$   & $20^{+4}_{-5}$   & $89^{+23}_{-43}$   & $121^{+3}_{-3}$   & $34^{+1}_{-1}$   \\ 
36399   & $20^{+12}_{-3}$   & $1008^{+32}_{-30}$   & $0.69^{+0.21}_{-0.10}$   & $101^{+541}_{-58}$   & $202^{+55}_{-115}$   & $113^{+166}_{-46}$   & $143^{+12}_{-20}$   & $39^{+1}_{-1}$   \\ 
37606   & $18^{+34}_{-8}$   & $383^{+9}_{-9}$   & $0.89^{+0.10}_{-0.38}$   & $97^{+45}_{-40}$   & $201^{+10}_{-176}$   & $87^{+176}_{-11}$   & $96^{+12}_{-6}$   & $43^{+2}_{-2}$   \\ 
38018   & $18^{+1}_{-1}$   & $556^{+8}_{-9}$   & $0.44^{+0.10}_{-0.11}$   & $509^{+34}_{-460}$   & $265^{+9}_{-175}$   & $41^{+172}_{-15}$   & $48^{+6}_{-7}$   & $33^{+1}_{-1}$   \\ 
38146   & $12^{+2}_{-1}$   & $889^{+22}_{-26}$   & $0.57^{+0.16}_{-0.13}$   & $522^{+69}_{-88}$   & $190^{+26}_{-169}$   & $151^{+153}_{-50}$   & $44^{+10}_{-11}$   & $11^{+1}_{-1}$   \\ 
38596   & $27^{+35}_{-9}$   & $2566^{+2582}_{-918}$   & $0.68^{+0.27}_{-0.20}$   & $1298^{+1905}_{-903}$   & $288^{+18}_{-170}$   & $99^{+157}_{-41}$   & $62^{+18}_{-13}$   & $31^{+1}_{-1}$   \\ 
38910   & $72^{+84}_{-29}$   & $3437^{+3199}_{-1354}$   & $0.32^{+0.50}_{-0.22}$   & $2063^{+3008}_{-1480}$   & $164^{+169}_{-109}$   & $117^{+60}_{-81}$   & $53^{+21}_{-14}$   & $51^{+2}_{-1}$   \\ 
38980   & $15^{+26}_{-4}$   & $991^{+111}_{-136}$   & $0.74^{+0.25}_{-0.37}$   & $439^{+247}_{-300}$   & $295^{+11}_{-176}$   & $105^{+164}_{-32}$   & $76^{+8}_{-8}$   & $34^{+1}_{-1}$   \\ 
39681   & $21^{+11}_{-3}$   & $1579^{+534}_{-225}$   & $0.66^{+0.22}_{-0.21}$   & $1031^{+468}_{-813}$   & $60^{+10}_{-8}$   & $72^{+31}_{-20}$   & $72^{+6}_{-6}$   & $15^{+2}_{-2}$   \\ 
39903   & $30^{+1}_{-1}$   & $922^{+11}_{-11}$   & $0.14^{+0.05}_{-0.04}$   & $125^{+51}_{-53}$   & $339^{+16}_{-334}$   & $35^{+162}_{-21}$   & $32^{+4}_{-4}$   & $49^{+1}_{-1}$   \\ 
40015   & $7^{+18}_{-1}$   & $245^{+5}_{-6}$   & $0.56^{+0.43}_{-0.27}$   & $196^{+32}_{-70}$   & $107^{+177}_{-33}$   & $116^{+51}_{-45}$   & $112^{+15}_{-12}$   & $33^{+1}_{-1}$   \\ 
42916   & $23^{+3}_{-2}$   & $833^{+15}_{-11}$   & $0.69^{+0.08}_{-0.08}$   & $384^{+36}_{-31}$   & $142^{+84}_{-77}$   & $104^{+71}_{-75}$   & $150^{+9}_{-10}$   & $36^{+1}_{-1}$   \\ 43352   & $9^{+2}_{-1}$   & $1174^{+187}_{-113}$   & $0.51^{+0.19}_{-0.14}$   & $451^{+488}_{-264}$   & $208^{+25}_{-161}$   & $125^{+156}_{-44}$   & $138^{+11}_{-9}$   & $13^{+1}_{-1}$   \\ 
49546   & $11^{+18}_{-3}$   & $566^{+34}_{-23}$   & $0.78^{+0.21}_{-0.25}$   & $466^{+72}_{-104}$   & $328^{+11}_{-174}$   & $79^{+165}_{-26}$   & $84^{+5}_{-8}$   & $16^{+1}_{-1}$   \\ 
50180   & $35^{+18}_{-4}$   & $1838^{+1754}_{-410}$   & $0.37^{+0.31}_{-0.18}$   & $1041^{+820}_{-680}$   & $194^{+5}_{-179}$   & $159^{+150}_{-101}$   & $77^{+4}_{-5}$   & $30^{+1}_{-1}$   \\ 
54424   & $10^{+2}_{-1}$   & $1124^{+270}_{-211}$   & $0.55^{+0.18}_{-0.17}$   & $615^{+438}_{-300}$   & $300^{+27}_{-159}$   & $126^{+165}_{-91}$   & $128^{+11}_{-10}$   & $15^{+1}_{-1}$   \\ 
54746   & $12^{+17}_{-2}$   & $1018^{+79}_{-50}$   & $0.83^{+0.16}_{-0.16}$   & $166^{+196}_{-86}$   & $85^{+16}_{-11}$   & $127^{+24}_{-25}$   & $70^{+8}_{-9}$   & $19^{+1}_{-1}$   \\ 
61100   & $31^{+5}_{-3}$   & $1420^{+303}_{-157}$   & $0.61^{+0.12}_{-0.10}$   & $201^{+817}_{-135}$   & $176^{+6}_{-14}$   & $51^{+32}_{-22}$   & $57^{+6}_{-6}$   & $41^{+1}_{-1}$   \\ 
62512   & $45^{+10}_{-3}$   & $1600^{+703}_{-190}$   & $0.20^{+0.27}_{-0.13}$   & $299^{+1066}_{-217}$   & $31^{+176}_{-2}$   & $164^{+10}_{-92}$   & $108^{+2}_{-2}$   & $39^{+1}_{-1}$   \\ 
73440   & $8^{+17}_{-2}$   & $448^{+13}_{-13}$   & $0.65^{+0.34}_{-0.26}$   & $109^{+66}_{-65}$   & $244^{+24}_{-163}$   & $84^{+156}_{-37}$   & $68^{+11}_{-10}$   & $30^{+1}_{-1}$   \\ 
75401   & $19^{+12}_{-3}$   & $963^{+707}_{-89}$   & $0.47^{+0.39}_{-0.29}$   & $677^{+292}_{-347}$   & $121^{+177}_{-11}$   & $132^{+39}_{-54}$   & $74^{+8}_{-9}$   & $19^{+1}_{-1}$   \\ 
76006   & $8^{+15}_{-2}$   & $545^{+18}_{-18}$   & $0.69^{+0.30}_{-0.33}$   & $220^{+68}_{-61}$   & $75^{+136}_{-35}$   & $111^{+61}_{-80}$   & $57^{+12}_{-12}$   & $14^{+1}_{-1}$   \\ 
78970   & $13^{+7}_{-1}$   & $871^{+76}_{-58}$   & $0.27^{+0.53}_{-0.16}$   & $663^{+183}_{-614}$   & $53^{+178}_{-14}$   & $100^{+66}_{-36}$   & $61^{+10}_{-7}$   & $22^{+2}_{-1}$   \\ 
80884   & $14^{+2}_{-1}$   & $1008^{+60}_{-39}$   & $0.46^{+0.17}_{-0.13}$   & $110^{+710}_{-80}$   & $291^{+8}_{-176}$   & $118^{+167}_{-22}$   & $64^{+6}_{-7}$   & $10^{+1}_{-1}$   \\ 
84062   & $11^{+2}_{-1}$   & $738^{+31}_{-46}$   & $0.39^{+0.28}_{-0.19}$   & $239^{+366}_{-188}$   & $27^{+15}_{-10}$   & $81^{+55}_{-48}$   & $60^{+8}_{-9}$   & $35^{+1}_{-1}$   \\ 
84223   & $30^{+50}_{-10}$   & $3607^{+5653}_{-1646}$   & $0.73^{+0.26}_{-0.32}$   & $1958^{+5637}_{-1470}$   & $151^{+176}_{-37}$   & $100^{+76}_{-45}$   & $110^{+16}_{-13}$   & $40^{+1}_{-1}$   \\ 
84949   & $15^{+9}_{-3}$   & $1604^{+565}_{-301}$   & $0.76^{+0.17}_{-0.17}$   & $451^{+766}_{-335}$   & $140^{+10}_{-15}$   & $62^{+18}_{-18}$   & $60^{+10}_{-10}$   & $15^{+1}_{-1}$   \\ 
85852   & $8^{+1}_{-1}$   & $903^{+48}_{-47}$   & $0.30^{+0.23}_{-0.16}$   & $640^{+178}_{-430}$   & $181^{+155}_{-51}$   & $111^{+99}_{-79}$   & $135^{+12}_{-10}$   & $11^{+1}_{-1}$   \\ 
88848   & $44^{+21}_{-8}$   & $2053^{+995}_{-462}$   & $0.73^{+0.16}_{-0.20}$   & $660^{+845}_{-460}$   & $253^{+4}_{-178}$   & $103^{+173}_{-12}$   & $78^{+3}_{-3}$   & $30^{+1}_{-1}$   \\ 
90355   & $11^{+8}_{-2}$   & $304^{+8}_{-11}$   & $0.40^{+0.48}_{-0.19}$   & $193^{+60}_{-58}$   & $155^{+53}_{-33}$   & $71^{+81}_{-47}$   & $130^{+15}_{-19}$   & $30^{+2}_{-2}$   \\ 
93137   & $13^{+7}_{-2}$   & $874^{+59}_{-52}$   & $0.61^{+0.29}_{-0.26}$   & $549^{+169}_{-134}$   & $28^{+5}_{-5}$   & $101^{+17}_{-39}$   & $89^{+5}_{-5}$   & $19^{+1}_{-1}$   \\ 
94347   & $13^{+1}_{-1}$   & $634^{+15}_{-15}$   & $0.24^{+0.13}_{-0.11}$   & $345^{+88}_{-70}$   & $32^{+150}_{-21}$   & $64^{+69}_{-36}$   & $146^{+10}_{-8}$   & $23^{+1}_{-1}$   \\ 
94802   & $18^{+1}_{-1}$   & $1037^{+35}_{-27}$   & $0.11^{+0.08}_{-0.06}$   & $559^{+184}_{-325}$   & $91^{+176}_{-7}$   & $102^{+59}_{-59}$   & $130^{+4}_{-4}$   & $24^{+1}_{-1}$   \\ 
97063   & $16^{+2}_{-1}$   & $1376^{+427}_{-124}$   & $0.25^{+0.32}_{-0.14}$   & $848^{+435}_{-585}$   & $48^{+175}_{-31}$   & $96^{+63}_{-61}$   & $36^{+10}_{-10}$   & $26^{+1}_{-1}$   \\ 
97690   & $9^{+2}_{-1}$   & $1063^{+62}_{-60}$   & $0.23^{+0.44}_{-0.13}$   & $387^{+436}_{-289}$   & $47^{+170}_{-16}$   & $109^{+58}_{-55}$   & $60^{+9}_{-9}$   & $10^{+1}_{-1}$   \\ 
98375   & $8^{+20}_{-2}$   & $284^{+10}_{-7}$   & $0.77^{+0.22}_{-0.40}$   & $177^{+61}_{-117}$   & $191^{+16}_{-170}$   & $85^{+165}_{-27}$   & $93^{+9}_{-5}$   & $20^{+1}_{-1}$   \\ 
103287   & $30^{+7}_{-2}$   & $1347^{+573}_{-188}$   & $0.35^{+0.38}_{-0.23}$   & $873^{+406}_{-661}$   & $339^{+5}_{-178}$   & $127^{+161}_{-46}$   & $88^{+6}_{-7}$   & $25^{+2}_{-2}$   \\ 
104440   & $88^{+106}_{-41}$   & $2429^{+3089}_{-1104}$   & $0.65^{+0.11}_{-0.07}$   & $1201^{+2500}_{-784}$   & $17^{+180}_{-3}$   & $147^{+32}_{-136}$   & $91^{+1}_{-1}$   & $51^{+1}_{-1}$   \\ 
105969   & $14^{+4}_{-2}$   & $867^{+41}_{-41}$   & $0.42^{+0.30}_{-0.22}$   & $314^{+386}_{-137}$   & $93^{+160}_{-14}$   & $110^{+38}_{-24}$   & $130^{+8}_{-8}$   & $15^{+3}_{-2}$   \\ 
107143   & $35^{+5}_{-2}$   & $1346^{+215}_{-86}$   & $0.29^{+0.23}_{-0.16}$   & $1092^{+213}_{-940}$   & $276^{+25}_{-175}$   & $106^{+156}_{-63}$   & $41^{+10}_{-10}$   & $40^{+2}_{-2}$   \\ 
110291   & $39^{+17}_{-26}$   & $794^{+60}_{-43}$   & $0.99^{+0.00}_{-0.19}$   & $562^{+177}_{-468}$   & $234^{+20}_{-168}$   & $87^{+175}_{-9}$   & $99^{+14}_{-4}$   & $15^{+3}_{-1}$   \\ 
113638   & $12^{+22}_{-3}$   & $366^{+11}_{-9}$   & $0.40^{+0.50}_{-0.26}$   & $186^{+71}_{-77}$   & $108^{+210}_{-89}$   & $111^{+59}_{-44}$   & $64^{+14}_{-12}$   & $31^{+7}_{-8}$   \\ 
117622   & $13^{+3}_{-1}$   & $1070^{+263}_{-70}$   & $0.44^{+0.23}_{-0.17}$   & $326^{+628}_{-248}$   & $33^{+7}_{-8}$   & $96^{+34}_{-37}$   & $119^{+9}_{-9}$   & $20^{+1}_{-1}$   \\ 
118212   & $38^{+6}_{-5}$   & $883^{+438}_{-28}$   & $0.58^{+0.18}_{-0.12}$   & $522^{+248}_{-90}$   & $112^{+169}_{-5}$   & $109^{+28}_{-23}$   & $113^{+4}_{-5}$   & $68^{+2}_{-3}$   \\ 

\enddata
\end{deluxetable}

\begin{deluxetable}{r|rrr|rrr|rrrc}
\tabletypesize{\scriptsize}
\rotate
\tablecaption{Comparison of astrometric fits with known spectroscopic orbits. \label{known.tab}}
\tablewidth{0pt}
\tablehead{
\multicolumn{1}{c}{}& \multicolumn{3}{c}{NDAC}  &  \multicolumn{3}{c}{FAST}  &  \multicolumn{4}{c}{Spectroscopic}  \\
\colhead{HIP} & \colhead{c}{$P$ [d]} & \colhead{$e$} & \multicolumn{1}{c}{$(M_2^3/M_{\rm tot}^2)$} &
\colhead{$P$ [d]} & \colhead{$e$} & \multicolumn{1}{c}{$(M_2^3/M_{\rm tot}^2)$} &
 \colhead{$P$ [d]} & \colhead{$e$} & \colhead{$f(M)$}  & \colhead{Ref.}}
\startdata
 12062 & $1375\pm 227$ &  $0.99_{-0.55}^{+0.01}$ & 5.16 &   $973\pm 140$ &  $0.64_{-0.28}^{+0.22}$  &  0.032 &   $905\pm 12$ &  $0.260\pm 0.032$  & $0.0262\pm 0.0038$  &   (1)   \\  
 19832 & $678\pm 27$ &  $0.23_{-0.17}^{+0.25}$ & 0.023 &   $668\pm 34$ &  $0.19_{-0.18}^{+0.37}$  &  0.023 &   $717\pm 3$ &  $0.074\pm 0.029$  &  &   (2)   \\  
 37606 & $384\pm 11$ &  $0.85_{-0.34}^{+0.08}$ & 0.056 &   $386\pm 10$ &  $0.85_{-0.39}^{+0.10}$  &  0.033 &   $380.6\pm 0.1$ &  $0.73\pm 0.012$  &  &   (3)   \\  
 38018 & $544\pm 11$ &  $0.48_{-0.11}^{+0.12}$ & 0.069 &   $555\pm 8$ &  $0.42_{-0.11}^{+0.12}$  &  0.068 &   551.9 &  0.39  &            &  (4) \\  
 40015 & $252\pm 7$ &  $0.99_{-0.56}^{+0.01}$ & 0.54   &   $246\pm 5$ &  $0.38_{-0.27}^{+0.44}$   & 0.010 &   243.8  &  0.42 &             &  (4) \\  
 61100 & $1284\pm 118$ &  $0.62_{-0.15}^{+0.23}$ & 0.046&  $1384\pm 227$ &  $0.59_{-0.09}^{+0.12}$ & 0.033 &   1284.4&  0.50 &            &   (3) \\  
 73440 & $435\pm 18$ &  $0.50_{-0.12}^{+0.14}$ & 0.0064&   $450\pm 13$&  $0.44_{-0.16}^{+0.34}$   & 0.0064 & $467.2\pm 9.7$  &  $0.2170\pm 0.077$ & $0.00078\pm 0.00018$  &   (1) \\  
 84949 & $89\pm 845$ &  $0.88_{-0.16}^{+0.09}$ & 47.6     & $1609\pm 462$ & $0.72_{-0.16}^{+0.16}$   & 0.041 &  $2018.8\pm 0.7$  & $0.672\pm 0.002$   &            &   (5)   \\  
 85852 & $916\pm 61$ &  $0.26_{-0.20}^{+0.30}$  & 0.069 &  $906\pm 46$ &  $0.18_{-0.14}^{+0.14}$   & 0.056 &  $903.8\pm 0.4$ & $0.072\pm 0.031$ &  $0.00351\pm 0.00031$ &   (6)          \\ 
 88848 & $4457\pm 776$ &  $0.95_{-0.21}^{+0.04}$ & 0.25  & $1460\pm 740$ & $0.43_{-0.20}^{+0.15}$   & 0.094 & $2092.2\pm 5.8$  & $0.765\pm 0.013$  & $0.113\pm 0.011$  &  (7)  \\  
105969& $887\pm 35$ & $0.23_{-0.18}^{+0.22}$ & $0.086$  &  $859\pm 41$ &  $0.32_{-0.21}^{+0.30}$   & 0.12   &   878    & 0.13  &            &  (8)  \\  
\enddata
\tablecomments{References: (1) \citet{lat02}; (2) \citet{ha}; (3) \citet{seti}; (4) \citet{ha03};
(5) \citet{sca}; (6) \citet{fek93}; (7) \citet{fek}; (8) \citet{pou}}
\end{deluxetable}


\clearpage

\begin{thebibliography}{}

\bibitem[Abt \& Levy(1985)]{abt} 
Abt, H.A., \& Levy, S.G., 1985, \apjs, 59, 229

\bibitem[Andrievsky \& Paunzen(2003)]{and} 
Andrievsky, S.M., \& Paunzen, S., 2003, \mnras, 313, 574

\bibitem[Carrier et al.(2002)]{car} 
Carrier, F., North, P., Udry, S., \& Babel, J., 2002, \aap, 394, 151

\bibitem[ESA(1997)]{esa} 
ESA, 1997, The Hipparcos and Tycho Catalogues, ESA SP-1200

\bibitem[Fabricius \& Makarov(2000)]{fama} 
Fabricius, C., \& Makarov, V.V., 2003, AN, 324, 419

\bibitem[Falin \& Mignard(1999)]{fal} 
Falin, J.L., \& Mignard, F., 1999, \aaps, 135, 231

\bibitem[Faraggiana et al.(2004)]{far04} 
Faraggiana, R., et al. 2004, \aap, 425, 615

\bibitem[Faraggiana \& Bonifacio(2005)]{far05} 
Faraggiana, R., \& Bonifacio, P., 2005, \aap, 436, 697

\bibitem[Fekel et al.(1993)]{fek93} 
Fekel, F.C., Henry, G.W., Busby, M.R., Eitter, J.J., 1993, \aj, 106, 2370

\bibitem[Fekel et al.(2005)]{fek} 
Fekel, F.C., et al., 2005, \aj, 129, 1001

\bibitem[Freire Ferrero et al.(2004)]{fre} 
Freire Ferrero, R., Frasca, A., Marilli, E., \& Catalano, S., 2004, \aap, 413, 657

\bibitem[Grenier et al.(1999)]{gre} 
Grenier, et al., 1999, \aaps, 135, 503

\bibitem[Halbwachs et al.(2000)]{ha} 
Halbwachs, J.L., Arenou, F., Mayor, M., Udry, S. \& Queloz, D., 2000, \aap, 355, 581

\bibitem[Halbwachs(2003)]{ha03} 
Halbwachs, J.L., Mayor, M., Udry, S., \& Arenou, F.,  2003, \aap, 397, 159

\bibitem[van Hamme et al.(1994)]{vanh} 
van Hamme, W.V., et al., 1994, \aj, 107, 1521

\bibitem[Heintz(1978)]{hei} 
Heintz, W.D., 1978, Double Stars, D. Reidel Publishing Company

\bibitem[Kaplan \& Makarov(2003)]{kama} 
Kaplan, G.H., \& Makarov, V.V., 2003, AN, 324, 419


\bibitem[Latham et al.(2002)]{lat02} 
Latham, D.W., et al., 2002, \aj, 124, 1144

\bibitem[Liu et al.(2002)]{liu} 
Liu, M.C., et al., 2002, \apj, 571, 519

\bibitem[Makarov \& Kaplan(2005)]{maka} 
Makarov, V.V., \& Kaplan, G,H.,  2005, \aj, 129, 2420

\bibitem[de Medeiros \& Mayor(1999)]{demed} 
de Medeiros, J.R.,  \& Mayor, M.,  1999, \aaps, 139, 433

\bibitem[de Medeiros et al(2002)]{dem02} 
de Medeiros, J.R., da Silva, J.R.P., \& Maia, M.R.G.,  2002, \apj, 578, 943

\bibitem[Nidever et al.(2002)]{nide} 
Nidever, D.L., et al., 2002, \apjs, 141, 503

\bibitem[Nordstr{\"o}m et al.(2004)]{nord} 
Nordstr{\"o}m et al., 2004, \aap, 418, 989

\bibitem[Pourbaix \& Jorissen(2000)]{pou} 
Pourbaix, D., \& Jorissen, A., 2000, \aaps, 145, 161

\bibitem[Scarfe et al.(1994)]{sca} 
Scarfe, C.D., et al., 1994, \aj, 107, 1529

\bibitem[Setiawan et al.(2004)]{seti} 
Setiawan, J., et al., 2004, \aap, 421, 241

\bibitem[Storn \& Price(1995)]{DE95}
Storn, R., Price, K., 1995, ``Differential Evolution - a simple and efficient adaptive scheme for global optimization over continuous spaces'',  International Computer Science Insitute Technical Report tr-95-012, Berkeley, CA

\bibitem[Torres (2006)]{tor}
Torres, G., 2006, \aj, 131, 1022

\bibitem[Hastie (2001)]{Statlearn}
Hastie, T.,  Tibshirani, R.,  \& Friedman, J., 2001, ``The Elements of Statistical Learning. Data Mining, Inference and Prediction", 
Springer Series in Statistics
 \end{thebibliography}
\end{document}